\documentclass[review]{elsarticle}

\usepackage{lineno,hyperref}

\usepackage{array}
\usepackage{wrapfig}
\usepackage{amsmath, amsthm, amsfonts}
\usepackage{graphicx}
\usepackage{mathtools}
\usepackage{url}
\usepackage{array}
\usepackage{multirow}
\usepackage{caption}
\usepackage{subcaption}
\usepackage{color,soul}

\usepackage[flushleft]{threeparttable} 
\usepackage{booktabs,caption}

\usepackage{makecell}

\makeatletter

\journal{Expert Systems with Applications}

\biboptions{authoryear}

\begin{document}

\begin{frontmatter}
\title{Complex Network based Supervised Keyword Extractor}

\author[mainaddress]{Swagata Duari\corref{mycorrespondingauthor}}
\cortext[mycorrespondingauthor]{Corresponding author}
\ead{sduari@cs.du.ac.in}

\author[mainaddress]{Vasudha Bhatnagar}
\ead{vbhatnagar@cs.du.ac.in}
\address[mainaddress]{Department of Computer Science, University of Delhi, New Delhi 110007, India}

\begin{abstract}
In this paper, we present a supervised framework for automatic keyword extraction from single document. We model the text as complex network, and construct the feature set by extracting select node properties from it. Several node properties have been exploited by unsupervised, graph-based keyword extraction methods to discriminate keywords from non-keywords. We exploit the complex interplay of node properties to design a supervised keyword extraction method. 

The training set is created from the feature set by assigning a label to each candidate keyword depending on whether the candidate is listed as a gold-standard keyword or not. Since the number of keywords in a document is much less than non-keywords, the curated training set is naturally imbalanced. We train a binary classifier to predict keywords after balancing the training set.

The model is trained using two public datasets from scientific domain and tested using three unseen scientific corpora and one news corpus. Comparative study of the results with several recent keyword and keyphrase extraction methods establishes that the proposed method performs better in most cases. This substantiates our claim that graph-theoretic properties of words are effective discriminators between keywords and non-keywords. We support our argument by showing that the improved performance of the proposed method is statistically significant for all datasets. We also evaluate the effectiveness of the pre-trained model on Hindi and Assamese language documents. We observe that the model performs equally well for the cross-language text even though it was trained only on English language documents. This shows that the proposed method is independent of the domain, collection, and language of the training corpora.
\end{abstract}


\end{frontmatter}


\section{Introduction}
\label{sec:introduction}
  \textit{Keywords} are special words that are typically embedded in documents and provide a compact and precise representation of the document content. Author-specified keywords for research articles and blogs not only convey the topics that the document covers, but are also used by search engines and document databases to efficiently locate information. \textit{Keywords} are also used for categorizing and clustering stories in news industry, document summarization, and recommendations. Keywords can also aid in constructing titles for articles, assigning tags to blogs, and so on.
  
  Not all documents on the Web, however, are accompanied by keywords assigned by authors, in which case important and relevant terms have to be extracted from the document itself. Inundated with the massive volume of digital documents available on the Internet, it is in-feasible to manually extract keywords. Consequently, NLP researchers continually strive towards improving automated methods for \textit{keyword extraction}(KE). Keyphrase extraction is considered as an extension of the keyword extraction task \citep{mihalcea2004textrank,rousseau2015main}.

Arising from the problem of automated index generation \citep{luhn1957statistical}, earliest keyword extraction techniques used statistical methods \citep{ortuno2002keyword,herrera2008statistical}
, which begot the advantage of language and domain independence. With recent popularity of machine learning approaches, supervised and unsupervised methods for keyword extraction have been in forefront of the research arena \citep{boudin2013comparison,bulgarov2015comparison,mothe2018automatic}. Supervised learning methods basically \textit{identify} the keywords (or keyphrases) by modeling the problem as binary classification task, while unsupervised methods \textit{extract} keywords by quantifying and ranking the words' \textit{embedded-ness} in text.

Though enrichment of features in supervised approaches and growing sophistication in techniques have achieved enhanced performance, inadvertently the methods have promoted fixation for document structure, language, domain, and collection. Several state-of-the-art supervised algorithms for keyword extraction fail to accommodate a generic design because of one of the following three reasons. First, they require linguistic knowledge and hence are  dependent on the language tools (for example the works of \cite{hulth2003improved}, \cite{nguyen2007keyphrase}, \cite{zhang2008automatic}, and  \cite{chuang2012without}). These methods generate language-dependent features that are specific to the language of the training set. 

Second, most of the existing supervised algorithms are domain dependent \citep{nguyen2007keyphrase, kim2010semeval, caragea2014citation}. For example, citations-enhanced keyword extraction \citep{caragea2014citation} works only when citation information is available. Thus, such techniques work well for scientific domain, but are not suitable for a generic domain that contains texts from news articles, blog articles, meeting transcripts, etc. \cite{nguyen2007keyphrase} extracted keyphrases from scientific papers by enriching the feature set with morphological information found in scientific text, which is also an example of domain-dependent keyphrase extraction. 

Third, existing supervised KE methods are collection-dependent because they use statistical features that are derived from the document collection \citep{nguyen2007keyphrase,chuang2012without,caragea2014citation,sterckx2016supervised}. Frequency-based statistical features like tf-idf, positions of occurrence, etc. are collection sensitive, and they change drastically with a slight alteration of the training set. 

In addition to above three primary reasons, some algorithms require external information from sources like Wikipedia \citep{medelyan2009human,zhang2017mike} or expert knowledge in the form of label-distribution to incorporate hints (e.g. a noun word occurring in the title) \citep{gollapalli2017incorporating}. This leaves a research gap for generic keyword extractor that can be applied on any text without considering its language, domain, or corpora. Recognizing this gap, we investigate the  feasibility of designing a keyphrase prediction model that is domain-, language-, and collection-independent.

Graph-based unsupervised KE methods represent text as graph\footnote{Alternatively, complex network. We use the terms `graph' and `network' interchangeably.}, and rely on the node properties to discriminate between keywords and non-keywords \citep{mihalcea2004textrank,litvak2011degext, rousseau2015main, florescu2017position}. These methods process one document at a time and are autonomous, which makes them collection and domain agnostic. However, these methods are dependent on the language tools as they perform POS tagging\footnote{Part-of-speech tagging.} for identifying candidate keywords (nouns and adjectives) \citep{mihalcea2004textrank, litvak2011degext,rousseau2015main,tixier2016graph,florescu2017position}. Because of this reason, graph based KE methods are not pliable for texts in resource-poor languages. It is noteworthy that unsupervised methods often report lower performance as compared to their supervised counterparts.

In this research, we aim to bolster performance of supervised learning approach with the advantages of graph-based keyword extraction methods, sans the bias towards domain or collection underlying the training data. The idea is inspired by consistent success of graph-based KE methods, which are typically unsupervised and weak performers compared to their supervised counterparts. We build over the domain and collection independence of graph-based KE methods and use graph-based node properties as features to develop a supervised model with improved performance. Additionally, we eliminate the language dependency by using statistical properties to filter candidate keywords from the text. Specific contributions of our research are listed below. 
\begin{enumerate}
    \item We devise supervised learning approach for automatic keyword extraction using graph-theoretic feature set (Section \ref{sec:focus_and_approach} - \ref{subsec:dataset-generation}).
    
    \item We empirically validate our claim that the  method is domain-, and collection-independent (Sections \ref{sec:experimental_setup} and \ref{sec:results-and-discussion}).
    
    \item Post keyword extraction, we generate keyphrases from the predicted keywords and demonstrate that our method performs comparably with the state-of-the-art supervised keyphrase extraction approaches (Section \ref{subsec:keyphrase_model}). 
    
    \item We evaluate the performance of our proposed method on texts from two India languages to establish language independence of the model (Section \ref{subsec:result-indian-language}).
\end{enumerate}

We do not delve into sophisticated deep learning based methods due to the limited volume of training set we have, and time required for training the model. We proceed with classic and simple classifiers as a proof of concept, and believe that use of deep learning techniques will enhance the performance of the predictive model.

\section{Related Works}
\label{sec:related-works}
Existing supervised methods for automatic keyword extraction tackle the problem as a phrase-based binary classification task, where keyphrases ($n$-grams) are extracted from the documents \citep{turney1999learning,witten1999kea,hulth2003improved,nguyen2007keyphrase,caragea2014citation,sterckx2016supervised}. These methods first create a labelled training set by constructing features for candidate phrases (or words) in the text and designate each phrase as either positive (keywords) or negative (non-keywords) by consulting the associated gold-standard list. The training set thus created is used to induce a predictive model, which predicts word (or phrase) from unseen documents as keyword or non-keyword. Several algorithms for inducing a predictive model have been explored, including CRF and SVM \citep{zhang2008automatic}, Bagged decision tree \citep{medelyan2009human}, Na\"ive Bayes \citep{caragea2014citation,sterckx2016supervised}, etc. 

Since eliciting good quality features is crucial for performance of the trained model, feature construction is recognized as the focal task in creation of training set for supervised KE approaches. Wide variety of features have been proposed to obtain high quality training set for inducing well performing models, e.g., tf-idf, POS tags, n-gram features, etc. \cite{hulth2003improved} reported that adding certain linguistic knowledge (e.g., syntactic features) to the training set improves performance of the automatic keyword extractor as compared to relying only on statistics-based features such as, term frequency, n-grams, etc. \cite{nguyen2007keyphrase} used morphological features of text in the training set in addition to simple statistics-based features, and designed a keyword extractor for scientific articles. \cite{medelyan2009human} incorporated information from external sources like Wikipedia to improve the training set. In addition to these, structural features of the document \citep{lopez2010grisp}, knowledge about domain and collection \citep{nguyen2007keyphrase,caragea2014citation}, citation-information \citep{caragea2014citation}, incorporating expert knowledge \citep{gollapalli2017incorporating}, and multidimensional information \citep{zhang2017mike} are some popular methods for enriching the training set. 

Unsupervised KE techniques largely comprise graph-based methods, which transform the text into a graph (complex network) and use graph-theoretic properties to rank keywords. These methods are largely word-based (i.e. unigrams are extracted) \citep{rousseau2015main,tixier2016graph,duari2019scake}, with a few being phrase-based (i.e. $n$-grams are extracted) \citep{mihalcea2004textrank,florescu2017position}. Node properties like PageRank \citep{mihalcea2004textrank}, PageRank along with position of the word in text \citep{florescu2017position}, degree centrality \citep{litvak2011degext}, coreness \citep{rousseau2015main}, etc. have been studied extensively in the past. Network representation of the text leverages unsupervised keyword extraction methods because of their independence from the influence of domain of the document or corpus. We aim to overcome the domain and collection dependence of supervised KE methods by using graph-based node properties as features for training. Furthermore, we also overcome the problem of language dependency by using a statistical filter for candidate selection while maintaining the efficacy of supervised KE methods.

\section{Methodology}
\label{sec:focus_and_approach}
Graph-based approaches for keyword extraction established that keywords possess certain properties, which impart special character to them. We hypothesize that succinct properties of keywords are revealed when the text is presented as graph. These properties are effective signals to discriminate between keywords and non-keywords. Accordingly, we employ node properties in the graph representation of text as features to fortify against dependence on linguistic, domain, collection, or structural features of the document. We propose a supervised framework to extract keywords from single document, which consists of the following steps.
\begin{enumerate}
    \item \label{step1}Select candidate keywords from each document, and construct the corresponding graph-of-text (Section \ref{subsec:modeling-text-graph}).
    
    \item Extract select node properties as features from each graph-of-text (Section \ref{subsec:properties_of_keywords}).
    
    \item Prepare the training set and induce a predictive model (Section \ref{subsec:dataset-generation}).

\end{enumerate}
Steps (i) and (ii) harbour innovative approaches that are detailed below. We use the model induced in step (iii) to predict keywords from unseen documents.
\section{Modeling Text as Complex Network}
\label{subsec:modeling-text-graph}
Text is modeled as a complex system, where the basic units, i.e. words, interact among each other to bring out the ideas that the author intends to communicate. The interaction between words can be mapped using various relationships, such as statistical, semantic, syntactic, discourse, cognitive, etc. \citep{blanco2012graph}. The most frequently used relation for automatic KE systems is co-occurrence based statistical relation \citep{mihalcea2004textrank,litvak2011degext,rousseau2015main,tixier2016graph,florescu2017position}.

We use a parameter-free and language-agnostic approach for creating complex networks from text as proposed in our previous work \citep{duari2019scake}. The network representation of text is created by - (i)~selecting a subset of words from the text as candidates (Section \ref{subsec:select-candidate-keywords}) and (ii)~using these candidate keywords as nodes, and forging relationships between nodes to create edges (Section \ref{subsec:network-construction}). We briefly describe the method below.

\subsection{Selecting Candidate Keywords}
\label{subsec:select-candidate-keywords}
In order to reduce the search space for possible keywords, we first eliminate frequently used non-content bearing words from processing. To do this, we perform \textit{stopword removal} using a standard English stop words list\footnote{\url{http://www.lextek.com/manuals/onix/stopwords2.html}.}. For non-English texts, a custom-curated stopwords list can be adopted to suit the requirement. We then apply a filter to identify candidate keywords from the remaining words. We use $\sigma$-index~\citep{ortuno2002keyword} as a statistical filter to perform this task. 

The $\sigma$-index of a word computes normalized standard deviation of the word's spacing distribution in successive occurrences, with higher values of $\sigma$-index indicating higher term relevance~\citep{ortuno2002keyword}. We adopt Herrera and Pury's \citep{herrera2008statistical} implementation of $\sigma$-index, where the $\sigma$-index of a word $w$ in a document $D$ is defined as below. 

Let, $N = |D|$ be the document length, $n$ be the number of occurrences of $w$, and $p_i$ be the position of $i^{th}$ occurrence of $w$. Note that $p_0 = 0$ and $p_{n+1} = N+1$. Then $\sigma(w)$ is computed as:
\begin{equation}
\sigma(w) = \frac{s(w)}{\mu(w)} \text{,}
\end{equation}
where $\mu(w) = \frac{\sum_{i=0}^n(p_{i+1} - p_i)}{n+1} = \frac{N+1}{n+1}$ is the average distance between successive occurrences of $w$ and $s(w) = \sqrt{\frac{\sum_{i = 0}^{n}((p_{i+1}-p_i)-\mu(w))^2}{n-1}}$ is the standard deviation of word occurrences. We retain top-$33\%$ words ranked by $\sigma$-index as candidate keywords, which drastically reduces the search space to one-third of the original length. 

Conventional graph-based keyword extraction methods use part-of-speech taggers and select nouns and adjectives as candidate keywords using linguistic tools~\citep{mihalcea2004textrank,rousseau2015main,florescu2017position}. This makes these approaches dependent on the linguistic tools and inefficacious for resource-poor languages. The use of statistical filter, $\sigma$-index, in our proposed approach imparts language-independence to this phase, and thus makes the approach language agnostic.

Please note that the computation of $\sigma$-index requires a word to occur at least twice in the document. This does not conflict with our goal because a word that occurs exactly once is unlikely to be a keyword. Furthermore, as words in short texts do not occur frequently, we omit the computation of $\sigma$-index for documents with less than $100$ unique words excluding stopwords. In such situation, each word is considered as a candidate keyword.

\subsection{Network Construction}
\label{subsec:network-construction}
We model text as a weighted, undirected graph $G=(V, E, W)$, where $V$ is the set of vertices that comprises the candidate keywords, $E \in V \times V$ is the set of edges, and $W$ is the corresponding weighted adjacency matrix. We use the conventional relationship of ``co-occurrence" of words to define edges between the nodes. Two nodes (candidate words) are linked if they co-occur in a sliding window of user specified size~\citep{mihalcea2004textrank,rousseau2015main}. In order to eliminate the user parameter (window size), we slide the window over two consecutive sentences \citep{duari2019scake}. This strategy begets advantages of capturing coherence in the flow of ideas that a sentence carries from its previous sentence. The links are weighted by the number of times the adjacent nodes (words) co-occur in the original text, and isolated nodes are ignored. 

\begin{figure}[!h]
\centering
\begin{subfigure}{.5\columnwidth}
\centering
\includegraphics[scale = 0.45]{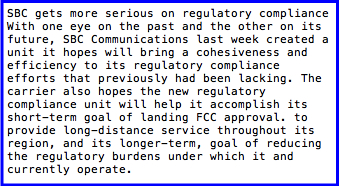}
\caption{\label{fig:text}Sample text.}
\end{subfigure}%
\begin{subfigure}{.5\columnwidth}
\centering
\includegraphics[scale = 0.5]{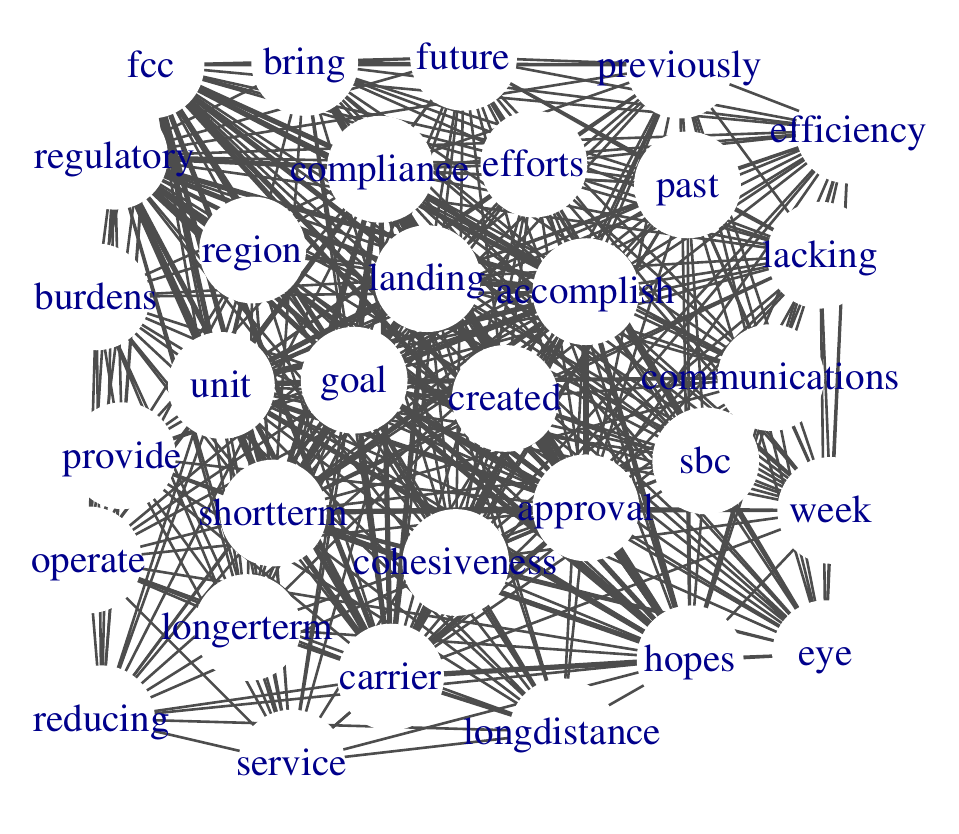}
\caption{\label{fig:network}Complex network of text.}
\end{subfigure}
\caption{\label{fig:network-for-text}Network created from sample text in Figure \ref{fig:text} (document id 6 from Hulth2003 dataset).}
\end{figure}

It is noteworthy that short texts (1-3 sentences) result into highly dense networks which are often complete graphs. Network density decreases as the number of sentences increases. Figure \ref{fig:network-for-text} shows network of a short text containing three sentences.

\section{Extracting Properties of Keywords}
\label{subsec:properties_of_keywords}
Centrality of nodes in a network is a popular estimate of node importance. According to \cite{boudin2013comparison}, degree centrality is conceptually the simplest and computationally most efficient centrality measure. However, in the context of weighted graph-of-text, it is more appropriate to use weighted degree (strength) as a measure of node importance \citep{barrat2004architecture}. Strength in our setting captures how often the words co-occur with each other in adjacent sentences.

\begin{figure}[!h]
    \centering
    \begin{subfigure}{.5\columnwidth}
        \includegraphics[scale = .23]{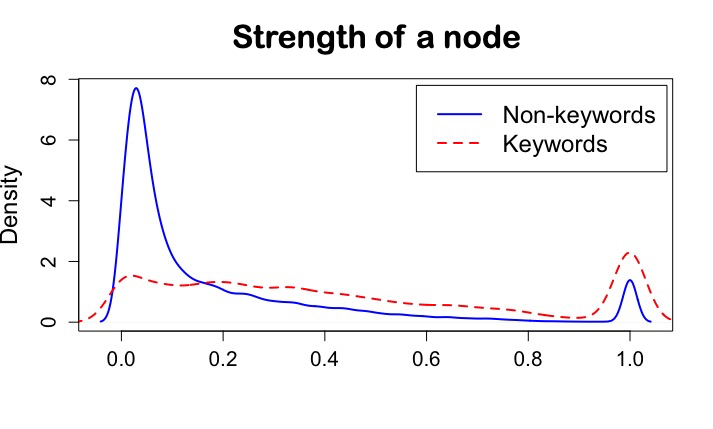}
        \caption{}
        \label{subfig:deg_densityl}
    \end{subfigure}%
    \begin{subfigure}{.5\columnwidth}
        \includegraphics[scale = .23]{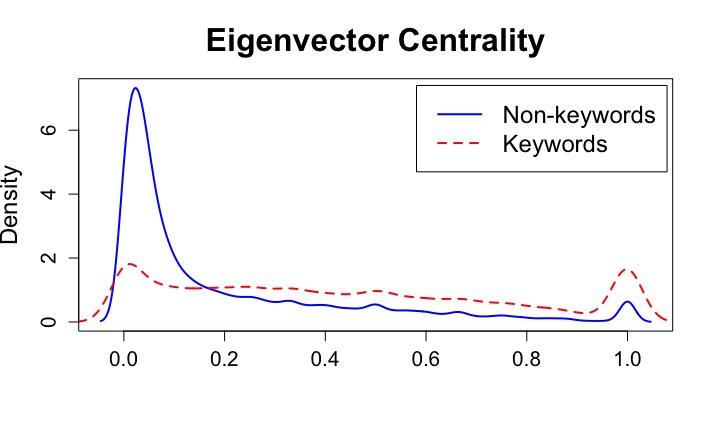}
        \caption{}
        \label{subfig:ev_densityl}
    \end{subfigure}
    \begin{subfigure}{.5\columnwidth}
        \includegraphics[scale = .23]{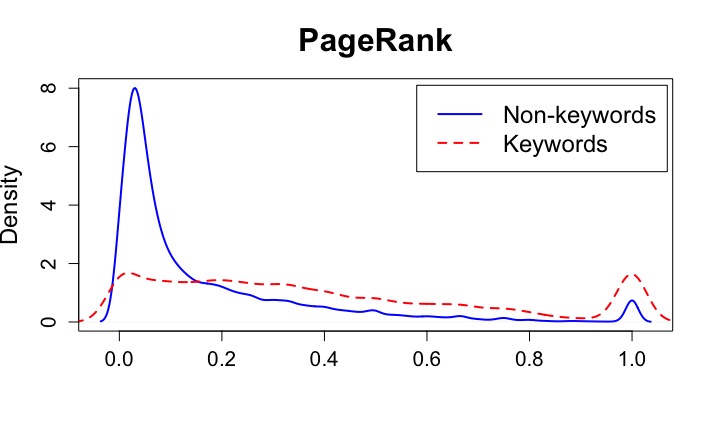}
        \caption{}
        \label{subfig:pr_densityl}
    \end{subfigure}%
    \begin{subfigure}{.5\columnwidth}
        \includegraphics[scale = .23]{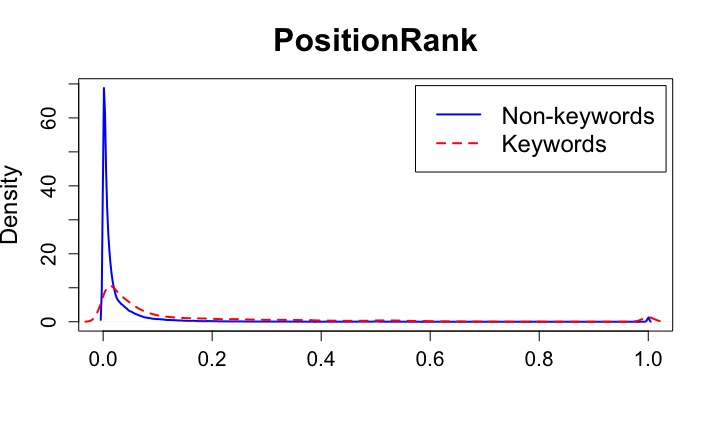}
        \caption{}
        \label{subfig:pos_densityl}
    \end{subfigure}
    \begin{subfigure}{.5\columnwidth}
        \includegraphics[scale = .23]{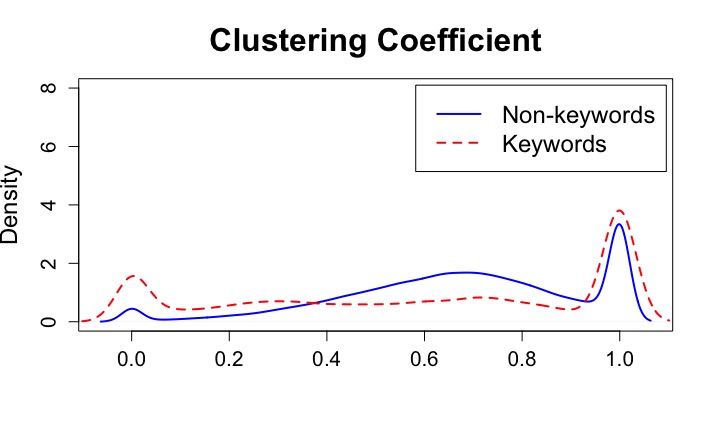}
        \caption{}
        \label{subfig:cc_densityl}
    \end{subfigure}%
    \begin{subfigure}{.5\columnwidth}
        \includegraphics[scale = .23]{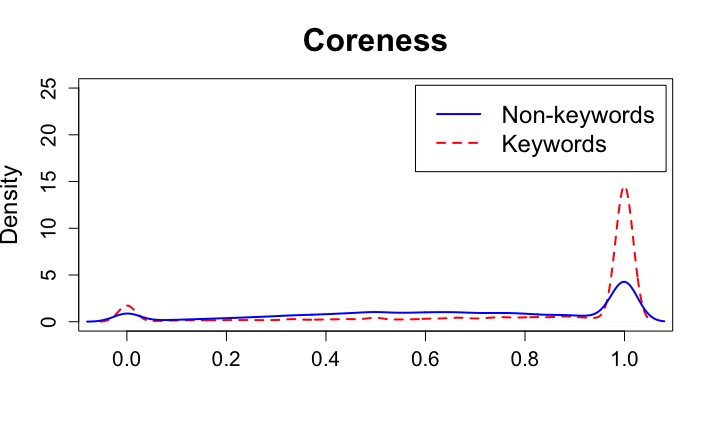}
        \caption{}
        \label{subfig:core_densityl}
    \end{subfigure}
    \caption{Density distributions of graph node properties for keywords and non-keywords using the SMOTE-balanced training set.}
    \label{Fig:density_distributions}
\end{figure}

Though strength effectively captures node importance, however, probability density distribution of strength for keywords and non-keywords for the training set prepared during our study clearly shows overlapping areas near high strength values (Figure \ref{subfig:deg_densityl}). The overlap indicates that strength alone is not an accurate discriminator between keywords and non-keywords. Next two plots (Figures \ref{subfig:ev_densityl} and \ref{subfig:pr_densityl}) show similar observations for Eigenvector centrality and PageRank. This impels exploration of other node properties - Coreness, PositionRank, and Clustering Coefficient - which would aid improvement in the quality of extracted keywords. It is noteworthy that we avoid centrality measures that require expensive all-pair shortest path computations. This maintains the frugality of feature extraction phase, enabling its potential for online usage.

All these properties, except Clustering Coefficient, have been individually vetted by state-of-the-art unsupervised graph-based keyword extraction methods. Our goal in this work is to investigate the complex interplay of these properties, which to the best of authors' knowledge, has not been explored for discriminating between keywords and non-keywords. We describe each of the node properties below.

\subsection{Strength of a node} 
Strength (weighted degree) of a node measures its \textit{embedded-ness} at local level. For node $v_i$ in a weighted network $G$, it is computed as~\citep{barrat2004architecture}:

$$strength(v_i) = \sum_j w_{ij} = \sum_j w_{ji}$$

Here, $w_{ij}$ is the corresponding entry in the weight matrix $W$ for edge $(v_i, v_j)$. High strength is associated with a node being more central or important in the network. The indulging intuition is that a word which is co-occurring with many other words (i.e., has high degree/strength) is important and is likely to be a keyword.

\subsection{Eigenvector centrality}
Eigenvector centrality or \textit{Prestige} of vertex $v_i$ quantifies its \textit{embedded-ness} in the network while recursively taking into account the prestige of its neighbors. Starting with initial prestige vector $\boldsymbol{\vec{p}_{0}}$ where all nodes (words) are assigned equal \textit{prestige}, $\boldsymbol{\vec{p}_{k}}$ is computed recursively as follows till convergence is achieved~\citep{zaki2014data}.
$$
\boldsymbol{\vec{p}_{k}} = W^T\boldsymbol{\vec{p}_{k-1}} = (W^k)^T\boldsymbol{\vec{p}_{0}}
$$

According to this computation, a word is important if it co-occurs with other important words. Nodes with higher eigenvector centrality are perceived as more important. Effectively, prestige of a word measures its influence over the entire document.

\subsection{PageRank} 
PageRank computes prestige in the context of `random surfer model' of Web search. A \textit{damping factor}, which is the probability of the surfer making \textit{random jump}, is used here. In case of text documents, this can be interpreted as events like the change of discourse in the document, beginning of a new chapter in a book, etc. We adopt the computation of word score ($WS$) from TextRank algorithm~\citep{mihalcea2004textrank}, as given below.

$$WS(v_i) = (1-d) + d * \sum_{v_j \in N_i} \left( \frac{w_{ji}}{\sum_{v_k \in N_j} w_{jk}} WS(v_j) \right)$$

Here, damping factor $d$ is set to 0.85 by the algorithm, which is the probability of random jump in context of the random surfing model. $N_i$ and $N_j$ are the sets of adjacent nodes of node $v_i$ and $v_j$, respectively. \cite{mihalcea2004textrank} expressed that the damping factor associated with random surfer model can be interpreted as text cohesion or ``knitting" of discourse together.

\subsection{PositionRank}
PositionRank is an extension of PageRank that is based on the intuition that keywords are likely to occur towards the beginning of the text rather than towards the end. PositionRank favors words occurring at the beginning of the document as keywords by using a position-biased weight for each candidate~\citep{florescu2017position}. Node $v_i \in V$ is assigned a weight based on its positional information by taking the inverse of the sum of its positions of occurrences in the text. Subsequently, PageRank computation is performed on the weighted nodes of the network to yield PositionRank scores for the candidate words. Formally, the PositionRank score of a node $v_i$ is computed as follows.

$$S(v_i) = (1 - \alpha).\Tilde{p_i} + \alpha.\sum_{v_j \in N_i}  \left( \frac{w_{ji}}{\sum_{v_k \in N_j} w_{jk}}S(v_j) \right)$$

Here, $\alpha$ is set as 0.85 by the algorithm, $\Tilde{p_i} = \frac{p_i}{\sum_{j = 1}^{|V|}p_j}$ is the normalized positional weight of $v_i$, $N_i$ is the set of adjacent nodes of $v_i$, and $w_{ji}$ is the weight of edge $e_{ji}$. The bias vector $\Tilde{p_i}$ ensures that words occurring towards the beginning are preferred as keywords by the system.

\subsection{Coreness} 
Coreness is a network degeneracy property that decomposes network $G$ into a set of \textit{maximal connected} subgraphs $G_k$ ($k$ denotes the core), such that nodes in $G_k$ have degree at least $k$ within the subgraph and $G_k \subseteq G_{k+1}$~\citep{seidman1983network}. \textit{Coreness} of a node is the highest core to which it belongs. \cite{rousseau2015main} presume that words in the main (highest) core of the network are keywords due to their dense connections. Though our findings differ where we have empirically observed that main core typically consists of fewer keyword-quality nodes that results in increasing precision and dropping recall \citep{duari2019scake}, we are convinced that keywords tend to lie in higher cores. Hence, we choose to include \textit{coreness} as a discriminating property.

\subsection{Clustering Coefficient} 
Clustering Coefficient (CC) of a node indicates edge density in its neighbourhood. It is a local property and is computed for node $v_i$ as the ratio of actual number of edges in the sub-graph induced by $v_i$ (excluding itself) to the total number of possible edges in that subgraph~\citep{zaki2014data}. A node $v_i$ having high clustering coefficient implies that the neighbors of $v_i$ are densely connected to each other, and can convey a context effectively without involving node $v_i$. For an undirected graph $G$, clustering coefficient of node $v_i \in G$ is computed as below.

$$CC(v_i) = \frac{2|{e_{jk}:v_j, v_k \in N_i, e_{jk} \in E}|}{|N_i|(|N_i|-1)}$$

Here, $N_i$ is the set of nodes adjacent to $v_i$. We conjecture that nodes (words) with low clustering coefficient connect diverse contexts together, and thus are likely to be important words. We elaborate the idea below.

\begin{figure}[!h]
\centering
\begin{subfigure}{.45\columnwidth}
\centering
\includegraphics[scale = 0.3]{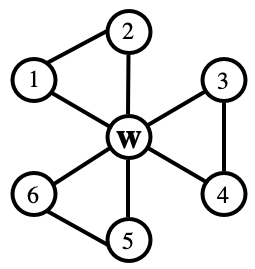}
\caption{\label{fig:unrel-context}Three semantically unrelated contexts, glued together by vertex $w$. CC for $w$ is 0.20 here.}
\end{subfigure}%
\hspace{.5cm}
\begin{subfigure}{.45\columnwidth}
\centering
\includegraphics[scale = 0.3]{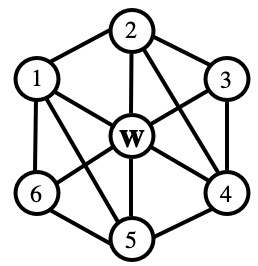}
\caption{\label{fig:rel-context}Semantically related contexts, causing higher clustering coefficient for vertex $w$. CC for $w$ is 0.53 here.}
\end{subfigure}
\caption{\label{fig:concepts}. Effect of semantically related and unrelated contexts on Clustering Coefficient.}
\end{figure}

A closed triad is formed in a graph of text when three words co-occur either in the same sentence or in adjacent sentences. Semantically, the words in the triad share a context. If a word $w$ shares many unrelated contexts with several words (Figure \ref{fig:unrel-context}), then $w$ attains importance because it glues several independent contexts. On the other hand, if the contexts in which $w$ participates are linked as shown in Figure \ref{fig:rel-context} (e.g., contexts formed by vertices 1,2,3 and vertices 1,4,5 are connected via an edge between vertices 2 and 4), then the word may not be important.

For unweighted networks, the above definition of topological clustering coefficient (CC) holds. However, for weighted networks, \cite{barrat2004architecture} define weighted clustering coefficient (WCC) that incorporates edge weights into computation. Since our networks are weighted, we empirically evaluated the effect of WCC against CC in distinguishing keywords from non-keywords. However, though WCC is apparently a better option, in our case the performance of the models degraded when using WCC in place of CC. This is because incorporating edge weights sometimes increases the clustering coefficient for the node, which negatively correlates with the importance of the node. Thus, we decided to use topological CC instead of WCC as a network property in our experiments.

Overlapping of densities of the six node property values in Figure \ref{Fig:density_distributions} indicate high number of false positives and likely poor performance. However, an intricate coaction of all six properties produces desirable effect of improving prediction performance, which has been established in Section \ref{sec:results-and-discussion}.

\section{Inducing the Model}
\label{subsec:dataset-generation}
The construction of training set is guided by our conjecture that the distribution of network-centric properties of the keywords are similar irrespective of the language, domain, or collection of the document. Accordingly, we combine  short scientific abstracts from Hulth2003 dataset and long scientific articles from SemEval2010 collection to create the training collection. The objective is to predict keywords from documents belonging to different collections of scientific papers, news articles, and non-English texts using the same predictive model.  

For each document in the training collection, we consult the corresponding gold-standard keywords list and assign the class label as `positive' or `negative' to the candidate words depending on whether they are listed as a gold-standard keyword (as unigram) or not. Corresponding feature values for the candidate keywords, as described in Section \ref{subsec:properties_of_keywords}, are range normalized to remove the bias due to document length. The feature set along with the labels constitute the training set for our empirical evaluations.

The curated training set naturally has high imbalance of class distribution because keywords are much lesser in number than other words. Longer documents in the training set contribute more to imbalance than shorter ones. Our pragmatism of using judicious mix of long and short text paid-off, resulting into the training set exhibiting an imbalance ratio of 1:5 (keywords:non keywords). We use Weka implementation of SMOTE~\citep{chawla2002smote} to balance\footnote{We set `percentage' parameter to $200\%$ for SMOTE filter.} the training set.

We use Na\"ive Bayes (NB) and XGBoost classifiers to train the model following their success as reported in earlier works. NB has been used for predicting keywords and keyphrases in various earlier studies \citep{nguyen2007keyphrase,medelyan2009human,kim2010semeval,caragea2014citation}. We decided to use Na\"ive Bayes classifier because of its simplicity, interpretability, and fast execution time. We additionally use a gradient boosted decision tree implemented in the XGBoost package (XGBoost classifier) following its success in predicting keyphrases as reported by Sterckx et al., who note that XGBoost classifier outperforms NB and linear classifiers in their study \citep{sterckx2016supervised}.

We attempt to reduce the high bias factor of NB classifier by balancing the training set using artificially generated data to over-sample the minority class. We additionally experiment with Bagging and Boosting ensembles of NB classifier to inspect for improvement in performance due to ensemble learning. Use of classical classifiers shows a marked improvement in performance in our experiments. We envisage that the performance will be further boosted by use of deep learning methods if sufficiently large dataset and efficient computation power is available.

\section{Experimental Setup and Objectives}
\label{sec:experimental_setup}
The proposed framework is implemented using R (version 3.3.1) and relevant packages\footnote{\url{https://cran.r-project.org/web/packages/}} (\texttt{igraph, tm, RWeka, caret and pROC}). We use six publicly available collections that have been used in similar studies. Each document in these collections is accompanied by an associated gold-standard keywords list, which is used as ground truth for testing the classifier performance. In the following sections, we briefly describe the datasets used in this study (Section \ref{subsec:dataset}) and the objective and design of our experiments (Section \ref{subsec:objective-experiment-design}).

\subsection{Datasets}
\label{subsec:dataset}

We use six publicly available datasets in our experiments. The datasets are described in detail below.
\begin{enumerate}
    \item Hulth2003 \citep{hulth2003improved} consists of 2000 scientific abstracts from \textit{Inspec} dataset, which are further divided into training (1000 articles), test (500 articles), and validation (500 articles). Each article in Hulth2003 dataset is accompanied with two gold-standard lists - one is controlled and uses a thesaurus, and the other is uncontrolled. We combine the training and test collections from Hulth2003 (a total of 1500 articles) to form a part of the training set for our experiments, and consult the uncontrolled keywords list as a gold-standard.
    
    \item WWW and KDD \citep{caragea2014citation} are two collections of abstracts from computer science articles published in the two well known conferences by the respective names. For both these collections, we consider only those articles which contains at least two sentences, and are accompanied by at least one gold-standard keyword. 
    
    \item Marujo2012 is a collection of 500 online news articles, which is grouped under training collection (450 articles) and test collection (50 articles). Each article is accompanied by a list of keywords assigned by human annotators through a HIT in Amazon Mechanical Turk \citep{marujo2012supervised}. From this dataset, we use the articles under training collection (a set of 450 articles) as an \textit{unseen test set} for our experiments.
    
    \item Krapivin2009 \citep{krapivin2009large} and SemEval2010 \citep{kim2010semeval} are two datasets which contains full scientific articles from ACM. The Krapivin2009 dataset consists of 2304 articles and associated keywords lists. SemEval2010 consists of 284 articles, out of which 144 are grouped as training, 100 as test, and 40 as validation sets. Each document in SemEval2010 dataset is accompanied by three sets of keyword list - author-assigned, reader-assigned, and author-and-reader-assigned. We use the combined collection of 244 articles (training and test) as a part of the training set for our experiments, and we consult the author-and-reader-assigned keywords list as gold-standard.
\end{enumerate}

Table \ref{tab:dataset_details} presents the datasets along with relevant statistics about the data. As mentioned in Section \ref{subsec:dataset-generation}, we create the training set for our experiments by combining the articles from Hulth2003 (1500 articles) and SemEval2010 (244 articles) datasets.

\begin{table*}[h!]
\scriptsize
\caption{\label{tab:dataset_details}Overview of the experimental data collections. $|D|$: Number of docs, $L_{avg}$: average doc length, $N_{avg}$: average gold-standard keywords per doc, $K_{avg}$: average gold-standard keyphrases per doc, $KP_{avg}$: average percentage of keywords present in the text, $ngram\%$: average \%age of 1/2/3/3+ -gram distribution.}
\begin{minipage}{\linewidth}
\begin{center}
\begin{tabular}{ l c c c c c c }
\hline
\textbf{Collection} & $\mathbf{|D|}$ & $\mathbf{L_{avg}}$ & $\mathbf{N_{avg}}$ & $\mathbf{K_{avg}}$ & $\mathbf{KP_{avg}}$ & \textbf{ngram\%}\\
\hline
Hulth2003 \citep{hulth2003improved} & 1500 & 129 & 23 & 10 & 90.07 & 16/52/24/8 \\
WWW \citep{caragea2014citation} & 1248 & 174 & 9 & 5 & 64.97 & 31/51/16/1 \\ 
KDD \citep{caragea2014citation} & 704 & 204 & 8 & 4 & 68.12 & 25/58/16/1 \\ 
Marujo2012 \citep{marujo2012supervised} & 450 & 427 & 69 & 48 & 99.31 & 75/17/5/2 \\ 
Krapivin2009 \citep{krapivin2009large} & 2304 & 7961 & 11 & 5 & 96.91 & 19/63/16/2 \\ 
SemEval2010 \citep{kim2010semeval} & 244 & 8085 & 34 & 16 & 95.89 & 21/55/20/4 \\ 
\hline
\end{tabular}
\end{center}
\end{minipage}
\end{table*}

\subsection{Objectives and Experimental design}
\label{subsec:objective-experiment-design}
We perform experimental evaluations to answer the following research questions.

\begin{enumerate}
    \item \label{exp:exp1}\textit{Which model performs best for automatic keyword extraction task?}\\ 
    We perform 10-fold cross-validation on the training set using XGBoost, Na\"ive Bayes, and Bagging and Adaboost ensembles of Na\"ive Bayes. To reduce the bias, we balance the training set using Weka implementation of SMOTE filter with percentage parameter set to $200\%$. Details of experiment and results are discussed in Section \ref{subsec:best-model}.
    
    \item \label{exp:exp2}\textit{How well do the graph-theoretic properties discriminate between the keywords and non-keywords over cross-collection and cross-domain datasets?}\\
    We use the best model trained in experiment \ref{exp:exp1} for all subsequent experiments. We perform cross-collection and cross-domain analysis of the trained model using three scientific datasets and one news corpus. Experimental results are discussed in Section \ref{subsec:domain-and-collection-independence}.
    
    \item \label{exp:exp3}\textit{How does the quality of extracted keyphrases compare with those extracted by state-of-the-art supervised and unsupervised keyphrase extraction methods?}\\
    We generate keyphrases from predicted keywords as a post processing step, and compare the quality with those extracted by state-of-the-art supervised and unsupervised keyphrase extraction methods. Comparative evaluation of the methods are presented for each dataset in Section \ref{subsec:keyphrase_model}.
    
    \item \label{exp:exp4}\textit{How well does the model perform on cross-language documents?}\\
    To substantiate our claim of the model being language-independent, we use the model trained in experiment \ref{exp:exp1} to extract keywords from documents written in Indian languages. Section \ref{subsec:result-indian-language} presents a detailed discussion on this experiment.
    
\end{enumerate}

\noindent
\paragraph{Evaluation Metrics}  We use precision, recall, and F1-score as performance evaluation metrics for all experiments pertaining to the above research questions. All three metrics are widely used in literature \citep{hulth2003improved,mihalcea2004textrank,medelyan2009human,caragea2014citation,rousseau2015main,sterckx2016supervised,zhang2017mike}. Except for 10-fold cross-validation results in Table \ref{tab:results-and-discussion}, all results presented in subsequent sections are macro-averaged at the dataset level.

\section{Results and Discussion}
\label{sec:results-and-discussion}
In this section, we present the results for our experiments as highlighted in Section \ref{subsec:objective-experiment-design}. Each subsection is devoted to one task, and we support our claim with empirical evidences.

\subsection{Establishing the Best Performing Model}
\label{subsec:best-model}
We trained four models on the SMOTE balanced training set each, using XGBoost \citep{chen2016xgboost}, Na\"ive Bayes (NB), and Bagging and Adaboost ensembles of NB. We present 10-fold cross validation results in Table \ref{tab:results-and-discussion}. Bold values represent best performance across all models in terms of the `positive' class\footnote{Positive class is of interest to us, which represents keywords}, and values in italics represent second-best results for the same. Among all models, XGBoost trained on the balanced training set turns out to be the best model and Adaboost ensemble of NB is the second best.

\begin{table}[h!]
\scriptsize
\caption{\label{tab:results-and-discussion}Cross-validated classifier performance. XGB: model trained using XGBoost classifier, NB: model trained using NB classifier.}
\begin{minipage}{\columnwidth}
\begin{center}
\begin{tabular}{ c | c | c | c }
\hline
\textbf{Models} & \textbf{P} & \textbf{R} & \textbf{F1} \\
\hline
XGB & \textbf{75.39} & \textbf{79.93} & \textbf{77.59} \\
NB & 72.4 & 49.71 & 58.95 \\
NB+Bagging (NB-B) & \textit{72.41} & 49.71 & 58.95 \\
NB+Adaboost (NB-A) & 72.20 & \textit{53.42} & \textit{61.41} \\
\hline
\end{tabular}
\end{center}
\end{minipage}
\end{table}

Next, we test the performance of trained models on test sets from Hulth2003 and SemEval2010 collections. Table \ref{tab:results-and-discussion-Test1} shows macro-averaged results on the test sets. Bold values indicate best performance for corresponding test sets and values in italics indicate second-best results. We observe that XGB classifier performs best in terms of recall and F1-score for both test sets, whereas best performance in terms of precision is achieved by NB and NB-B classifiers. NB-A performs second-best in term of recall and F1-score, following XGB. Since the performance gap in precision between NB, NB-B, and NB-A are insignificant, we decided to retain NB-A model along with XGB for all further experiments.

\begin{table}[ht!]
\scriptsize
\begin{minipage}{\columnwidth}
\caption{\label{tab:results-and-discussion-Test1}Macro-averaged results of model performances on test sets. Models are trained on the SMOTE balanced training set.}
\end{minipage}
\begin{minipage}{\columnwidth}
\begin{center}
\begin{tabular}{ c | c | c | c | c | c | c }
\hline
\multirow{2}{*}{\textbf{\makecell{Models}}} & \multicolumn{3}{c|}{Hulth2003 Test} & \multicolumn{3}{c}{Semeval2010} \\
 \cline{2-7}
 & P & R & F1 & P & R & F1\\
 \hline
 XGB & 49.8 & \textbf{83.5} & \textbf{60.7} & \textit{46.2} & \textbf{49 }& \textbf{46.4} \\
 NB & \textbf{52.8 }& 60.6 & 50.1 & \textbf{46.4} & 36.5 & 39.5 \\
 NB-B & \textbf{52.8} & 60.5 & 50.1 & \textbf{46.4} & 36.5 & 39.5 \\
 NB-A & \textit{52.4} & \textit{63.5} & \textit{51.8} & 45.2 & \textit{39.2} & \textit{40.6} \\
\hline
\end{tabular}
\end{center}
\end{minipage} 
\end{table}

\subsection{Establishing Domain and Collection Independence}
\label{subsec:domain-and-collection-independence}
With an aim to validate the claim of domain- and collection-independence, we evaluate XGB and NB-A on three unseen scientific datasets, i.e. Krapivin2009, KDD, and WWW, and one news corpus, i.e. Marujo2012. Recall that both models are trained on combined datasets from Hulth2003 and Semeval2010. Sections \ref{subsubsec:cross-collection-result} and \ref{subsubsec:cross-domain-result} present our results for establishing collection- and domain-independence, respectively.

\subsubsection{Result on Cross-collection datasets}
\label{subsubsec:cross-collection-result}
Table \ref{tab:cross-collection-res} shows macro-averaged results for the models on the three cross-collection scientific datasets. We observe that the models are able to recall the keywords reasonably well from unseen documents across corpora. To the best of our knowledge, no earlier work on supervised KE has performed cross-collection investigation for keyword extraction. Hence we are unable to compare the performance.

\begin{table}[ht!]
\scriptsize
\begin{minipage}{\columnwidth}
\caption{\label{tab:cross-collection-res}Macro-averaged results for XGB and NB-A on unseen cross-collection datasets.}
\end{minipage}
\begin{minipage}{\columnwidth}
\begin{center}
\begin{tabular}{ c | c | c | c | c | c | c }
\hline
\multirow{2}{*}{\textbf{\makecell{Test Sets}}} & \multicolumn{3}{c|}{XGB} & \multicolumn{3}{c}{NB-A} \\
 \cline{2-7}
 & P & R & F1 & P & R & F1\\
 \hline
 Krapivin2009 & 21.6 & \textbf{66.5} & 30.9 & \textbf{26.2} & 61.7 & \textbf{34.9}\\
 WWW & 14 & \textbf{81.8} & 23.1 & \textbf{24} & 66.3 & \textbf{33.1}\\
 KDD & 13.6 & \textbf{78.1} & 22.3 & \textbf{24.3} & 70.6 & \textbf{33.8}\\
\hline
\end{tabular}
\end{center}
\end{minipage}
\end{table}

Low precision for these datasets is due to the relatively less number of gold-standard keywords assigned per document (See column $N_{avg}$ in Table \ref{tab:dataset_details}). The models recall most of these keywords along with some false positives, which drops precision. NB-A outperforms XGB for these three datasets in terms of precision and F1-score, however with lower values for recall.

\subsubsection{Result on Cross-domain dataset}
\label{subsubsec:cross-domain-result}
We perform experiments to establish domain-independence of the trained models by evaluating their performance on an unseen, cross-domain dataset of news articles (Marujo2012). We present our empirical observations in Table \ref{tab:cross-domain-res}. It is evident from the results that the models are able to perform sufficiently on the cross-domain dataset, which establishes that the models are indeed applicable on documents from any domain. Again, we can't compare with any baseline due to reason stated in Section \ref{subsubsec:cross-collection-result}.

\begin{table}[ht!]
\scriptsize
\begin{minipage}{\columnwidth}
\caption{\label{tab:cross-domain-res}Macro-averaged results for XGB and NB-A on unseen cross-domain datasets.}
\end{minipage}
\begin{minipage}{\columnwidth}
\begin{center}
\begin{tabular}{ c | c | c | c | c | c | c }
\hline
\multirow{2}{*}{\textbf{\makecell{Test Sets}}} & \multicolumn{3}{c|}{XGB} & \multicolumn{3}{c}{NB-A} \\
 \cline{2-7}
 & P & R & F1 & P & R & F1\\
 \hline
 Marujo2012 & 58.3 & \textbf{42} & \textbf{45.2} &\textbf{ 67.4} & 29.8 & 37\\
\hline
\end{tabular}
\end{center}
\end{minipage}
\end{table}

Interestingly, for the cross-domain dataset (i.e., Marujo2012), the models are able to extract keywords with high precision, albeit with a drop in recall. This is because of the relatively higher number of keywords assigned per document ($N_{avg} = 69$ in Table \ref{tab:dataset_details}) for this dataset. The models tend to extract fewer but correct keywords, thus increasing precision and lowering recall in this case. XGB outperforms NB-A for this dataset in terms of recall and F1-score, whereas NB-A reports better precision.

\subsection{Comparison with Keyphrase Extraction Algorithms}
\label{subsec:keyphrase_model}
State-of-the-art supervised KE methods are \textit{phrase}-based extractors, whereas the unsupervised graph-based methods are \textit{word}-based extractors. Several earlier works suggest that keyphrase extraction should be treated as an extension of keyword extraction, and not as a separate task \citep{mihalcea2004textrank,rousseau2015main,papagiannopoulou2018local}. Following this viewpoint, we generate candidate keyphrases from the text as a post-processing step considering only those words which are predicted as keywords by our model.

We pre-process the text to remove stopwords, and split at punctuation marks to get phrases. All unique phrases that are not sub-strings of other phrases are extracted as keyphrases. We apply stemming\footnote{This is an optional step and can be skipped if stemmer is not available.} using Porter stemmer\footnote{\url{http://snowball.tartarus.org/algorithms/porter/stemmer.html}} to both the gold-standard keyphrases and the extracted keyphrases to improve performance of the keyphrase extractor.

For all datasets except Marujo2012, we extract top-5, -10, and -15 keyphrases based on our observation in column $K_{avg}$ in Table \ref{tab:dataset_details}. For Marujo2012, we extract top-5 to top-30 keyphrases (in increments of 5) to account for the higher number of average keywords assigned per document. 

We compare the performance of our models with the best in literature for each dataset. We report our observations for each dataset separately, because (i) results of all methods are not easily reproducible as their implementations are not publicly available, (ii) there is a diversity in choice of datasets for which the authors base their claims, and (iii) all state-of-the-art methods are not applicable across domain and corpora. 

We present our results in subsequent sections (\ref{subsubsec:hulth-result}-\ref{subsubsec:marujo-result}), comparing best performance of our models with select state-of-the-art methods evaluated on the datasets that we are using. We briefly explain these methods in subsequent sections and present comparative evaluation in the form of Tables. We also test the statistical significance of the improved performance of our algorithms over the baselines for each dataset (Section \ref{subsubsec:statistical-significance}).

\subsubsection{Result on Hulth2003 Test dataset}
\label{subsubsec:hulth-result}
In this section, we evaluate the performance of our KE models on Hulth2003 dataset with the works of \cite{hulth2003improved} and \cite{mihalcea2004textrank}. Hulth2003 dataset was curated by \cite{hulth2003improved} for her study of effect of linguistic properties in improving performance of keyword extractors. Later, this dataset has been mostly used by unsupervised keyword extraction methods \citep{rousseau2015main, duari2019scake}.

Hulth's work is supervised machine learning based, which uses linguistic information to improve performance. The method explores three term selection strategies - $n$-gram, noun-phrase (NP) chunk, and POS tag sequence, and evaluates the model performance on feature sets with and without POS tag information. Best result is obtained on POS tag based feature sets in comparison to their counterparts, and best F1-score is obtained with $n$-gram approach with POS tags as features.

\cite{mihalcea2004textrank} proposed an unsupervised approach, called TextRank, to extract keywords. The method is based on graph representation of text, where nouns and adjectives constitute the vertices set, and edges are formed between two vertices if they co-occur within a window of size $w$. Edges are undirected, and are weighted by the co-occurrence frequency of the adjacent vertices (words). PageRank \citep{brin1998anatomy} computation is performed on the graph representation of text to rank the vertices in order of their keyword-ness, with high PageRank score associated with being more likely to be a keyword. The system then selects top one-third candidates as keywords.

We report our results in Table \ref{tab:hulth-res}. It is clearly evident from the table that both XGB and NB-A outperform the baseline methods with large margin, with XGB leading in terms of precision, recall, and F1-score. Best result for both these models is obtained when we extract top-10 keyphrases, and XGB dominates NB-A on this dataset. It is noteworthy that the number of extracted keyphrases, i.e., 10 for Hulth2003 dataset, corresponds to the average number of keyphrases for the dataset as presented in Column $K_{avg}$ of Table \ref{tab:dataset_details}.

\begin{table}[!h]
    \scriptsize
      \centering
        \caption{Performance evaluation for Keyphrase Extraction on Hulth2003 Test dataset. $@k$: evaluation results for top-$k$ keyphrases.}
        \label{tab:hulth-res}
        \begin{tabular}{ c | c | c | c }
        \hline
        \textbf{Model} & $\mathbf{P}$ & $\mathbf{R}$ & $\mathbf{F1}$\\
        \hline
        XGB@10 & \textbf{52.5} & \textbf{65.1} & \textbf{54.7} \\
        NB-A@10 & 49.3 & 60.6 & 51.1 \\
        $n$-gram w. tag \citep{hulth2003improved} & 25.2 & 51.7 & 33.9 \\
        TextRank \citep{mihalcea2004textrank} & 31.2 & 43.1 & 36.2 \\
        \hline
        \end{tabular}
\end{table}

\subsubsection{Result on KDD and WWW dataset}
\label{subsubsec:KDDnWWW-result}
KDD and WWW datasets were curated by \cite{caragea2014citation} to study the effectiveness of citation information in improving the keyword extraction task. Since the study by \cite{caragea2014citation} uses citation information, the method is inefficacious for generic documents outside academic or scientific literature that do not have citation information. We evaluate the performance of XGB and NB-A models on KDD and WWW datasets, and compare them with two supervised baselines - CeKE \citep{caragea2014citation} and MIKE \citep{zhang2017mike}.

As mentioned above, CeKE enhances the feature set by using citation information along with statistical (tf-idf, position of occurrence, etc.) and linguistic (part-of-speech tags) information. The approach uses Na\"ive Bayes classifier to build a predictive model to identify keywords. On the other hand, MIKE uses multidimensional information (e.g., topical information) to enhance the feature set. It uses gradient-descent algorithm to build the predictive model. 

Table \ref{tab:kddwww-res} shows that XGB and NB-A outperform the two baselines with large margins in terms of precision, recall and F1-score. Performance of both XGB and NB-A models is comparable for the two datasets, with no (statistically) significant difference in performance of the models. Specifically, NB-A performs best for KDD dataset when we extract top-5 keyphrases, and XGB performs best on WWW dataset for the same number of keyphrases. The number of extracted keyphrases for both these models (i.e., 5 in this case) corresponds to the average number of keyphrases for both these datasets (column $K_{avg}$ in Table \ref{tab:dataset_details}).

\begin{table}[!h]
    \scriptsize
      \centering
        \caption{Performance evaluation for Keyphrase Extraction on KDD and WWW datasets. $@k$: evaluation results for top-$k$ keyphrases.}
        \label{tab:kddwww-res}
        \begin{threeparttable}
        \begin{tabular}{ >{\centering\arraybackslash}p{4cm} | c | c | c | c | c | c }
        \hline
        \multirow{2}{*}{\textbf{Model}} & \multicolumn{3}{c|}{KDD} & \multicolumn{3}{c}{WWW} \\
        \cline{2-7}
        & $\mathbf{P}$ & $\mathbf{R}$ & $\mathbf{F1}$ & $\mathbf{P}$ & $\mathbf{R}$ & $\mathbf{F1}$\\
        \hline
        XGB@5 & 26.9 & 49.7 & 33.3 & \textbf{30.3} & \textbf{52.3} & \textbf{36.6} \\
        NB-A@5 & \textbf{27.5} & \textbf{50.9} & \textbf{34.1} & 30.3 & 52 & 36.5 \\
        MIKE@5 \citep{zhang2017mike} & 14.01 & 17.33 & 15.49 & 14.8 & 15.05 & 14.92\\
        CeKE\tnote{*} \citep{caragea2014citation} & 21.3 & 41.3 & 28.0 & 22.7 & 38.6 & 28.4 \\
        \hline
        \end{tabular}
        \begin{tablenotes}
            \item[*] Results are averaged at document-level for 10-fold cross validation.
        \end{tablenotes}
        \end{threeparttable}
    \end{table}

\subsubsection{Result on Krapivin2009 dataset}
\label{subsubsec:krapi-result}
We evaluate the performance of XGB and NB-A models on Krapivin2009 dataset, and compare with one unsupervised baseline. The unsupervised baseline is a recent work by \cite{papagiannopoulou2018local}, which uses GloVe to encode local word embeddings for the terms in title and abstract of a scientific publication. A mean \textit{reference} vector is computed from the vectors trained from the full-text, and keyphrases are extracted by ranking all terms on the basis of their cosine similarity to the reference vector. Reference vector represents the semantics of the complete document, and words closer to it are considered keyphrases. RVA (Reference Vector Algorithm from abstracts) with 50-dimensional vector representation reports best result in terms of F1-score.

We present our experimental results on Krapivin2009 dataset in Table \ref{tab:krapi-res}. We observe that RVA performs best for this dataset in terms of F1-score. This shows the effectiveness of word embeddings in determining keyphrases. Blank entries (`-') in the table mean unavailability of results in relevant literature.

\begin{table}[!h]
    \scriptsize
    \centering
    \caption{Performance evaluation for Keyphrase Extraction on Krapivin2009 dataset. $@k$: evaluation results for top-$k$ keyphrases.}
    \label{tab:krapi-res}
    \begin{threeparttable}
        \begin{tabular}{ >{\centering\arraybackslash}p{4cm} | >{\centering\arraybackslash}p{1cm} | >{\centering\arraybackslash}p{1cm} | >{\centering\arraybackslash}p{1cm} }
            \hline
            \textbf{Model} & $\mathbf{P}$ & $\mathbf{R}$ & $\mathbf{F1}$\\
            \hline
            XGB@5 & 28.1 & 29.8 & 27.7 \\
            NB-A@5 & 27.2 & 28.6 & 26.7 \\
            RVA\tnote{*} \citep{papagiannopoulou2018local} & - & - & \textbf{32.06} \\
            \hline
        \end{tabular}
        \begin{tablenotes}
            \item[*] The algorithm is evaluated for top one-third keyphrases.
        \end{tablenotes}
    \end{threeparttable}
\end{table}

However, it is noteworthy that the evaluation of the baseline and our models is not same. The baseline is evaluated for top one-third keyphrases, whereas our models are evaluated for top-5 predicted keyphrases. This makes it difficult for us to perform an unbiased comparison of the methods. Among XGB and NB models, XGB performs best when we extract top-5 keyphrases. The number of keywords extracted (i.e., 5 in this case) correlates with the average number of keyphrases per document for Krapivin2009 dataset ($K_{avg} = 5$ in Table \ref{tab:dataset_details}).

\subsubsection{Result on SemEval2010 dataset}
\label{subsubsec:semeval-result}
SemEval2010 dataset was curated for Task 5 of the Workshop for Semantic Evaluation, 2010. 21 teams participated in the task, and HUMB \citep{lopez2010humb} performed best for author-and-reader-assigned keywords \citep{kim2010semeval}. 

We compare our XGB and NB-A models with HUMB \citep{lopez2010humb} and Boudin's algorithm \citep{boudin2018unsupervised} as baselines. HUMB is a supervised method that identifies keyphrases using a predictive model trained on a feature set of document structure (e.g. section and position), content (e.g. tf-idf), and external information (GRISP terminology and Wikipedia). The model is initially trained using a bagged decision tree, and candidates are further re-ranked using a probabilistic model to improve their ranking \citep{lopez2010humb}. Boudin's algorithm is unsupervised, which uses a multipartite graph representation of the text to encode keyphrase candidates and topics in a single graph. Candidates are ranked using TextRank computation for weighted graphs.

Table \ref{tab:semeval-res} presents the experimental results for the proposed models and the two baselines. We observe that XGB model outperforms all models in terms of precision, recall, and F1-score when we extract top-10 keyphrases. We also show the results of our models for top-15 keyphrases ($K_{avg} = 16$ in Table \ref{tab:dataset_details} for SemEval2010 dataset). However, we only show results of one baseline, HUMB, as Boudin's algorithm do not report results for top-15 keyphrases. The difference in performance of our models (i.e., XGB and NB) for top-10 and top-15 keyphrases is insignificant, with a slightly better performance for top-10 keyphrases.

\begin{table}[!h]
    \scriptsize
    \centering
    \caption{Performance evaluation for Keyphrase Extraction on SemEval2010 dataset. $@k$: evaluation results for top-$k$ keyphrases.}
    \label{tab:semeval-res}
    \begin{tabular}{ c | c | c | c }
        \hline
        \textbf{Model} & $\mathbf{P}$ & $\mathbf{R}$ & $\mathbf{F1}$\\
        \hline
        XGB@10 & \textbf{38.5} & \textbf{25.6} & \textbf{30.3} \\
        NB-A@10 & 36 & 24 & 28.3 \\
        HUMB@10 \citep{lopez2010humb} & 32.0 & 21.8 & 26.0 \\
        Boudin@10 \citep{boudin2018unsupervised} & - & - & 14.5 \\
        \hline
        XGB@15 & 30 & 29.9 & 29.5 \\
        NB-A@15 & 28.6 & 28.4 & 28.1 \\
        HUMB@15 & 27.2 & 27.8 & 27.5 \\
        \hline
    \end{tabular}
\end{table}

\subsubsection{Result on Marujo2012 dataset}
\label{subsubsec:marujo-result}
Marujo2012 dataset \cite{marujo2012supervised} is a cross-domain dataset that we adopted to establish domain-independence of our proposed method. The dataset consists of news articles. To compare the performance of XGB and NB-A models, we consider as baseline Boudin's algorithm \citep{boudin2018unsupervised}, which has already been briefed in Section \ref{subsubsec:semeval-result}.

We present our experimental results in Table \ref{tab:marujo-res}. We observe that our models outperformed the baseline by a huge margin and shows impressive performance for a cross-domain keyphrase extraction model. Specifically, best precision is achieved when we extract top-10 keyphrases using NB-A model, and best recall and F1-score is achieved when we extract top-30 keyphrases using the XGB model. High precision and comparatively low recall is due to the high number of gold-standard keyphrases assigned for this dataset (Table \ref{tab:dataset_details}, column $K_{avg}$). Our models predicted lesser number of keyphrases as in the gold-standard list, out of which most are correctly extracted (high precision) but a few correct keyphrases are missed (low recall).

\begin{table}[!h]
    \scriptsize
    \centering
    \caption{Performance evaluation for Keyphrase Extraction on Marujo2010 dataset. $@k$: evaluation results for top-$k$ keyphrases. *: Evaluated only for top-5 and top-10 keyphrases.}
    \label{tab:marujo-res}
    \begin{tabular}{ c | c | c | c }
        \hline
        \textbf{Model} & $\mathbf{P}$ & $\mathbf{R}$ & $\mathbf{F1}$\\
        \hline
        XGB@30 & 83.4 & \textbf{43.1} & \textbf{53.8} \\
        NB-A@30 & 80.81 & 33.36 & 44.64 \\
        \hline
        XGB@10 & 92.86 & 25.62 & 38.33 \\
        NB-A@10 & \textbf{92.91} & 25.55 & 38.21 \\
        Boudin@10* \citep{boudin2018unsupervised} & - & - & 18.2 \\
        \hline
    \end{tabular}
\end{table}

\subsubsection{Statistical Significance Testing}
\label{subsubsec:statistical-significance}
Our next goal is to examine if the performance of our algorithm is (statistically) significantly better than that of the corresponding baselines for each dataset. Since we know only the macro-averaged metrics for the baselines, we can't use traditional statistical significance testing approaches. Therefore, we follow the approach recommended by \cite{berg2012empirical} and \cite{dror2018hitchhiker}.

Let $O$ be our algorithm and $B$ be the baseline algorithm. We test the null hypothesis, H0: the performance of $O$ is no better than the performance of $B$, against the alternative, H1: the performance of $O$ is significantly better than $B$. We compare our method with the corresponding baselines for each dataset. The performance difference, $\delta(x)$, is the difference in performance metric of $O$ minus $B$ for the dataset $x$.

\begin{figure}[!h]
    \centering
    \includegraphics[scale = .4]{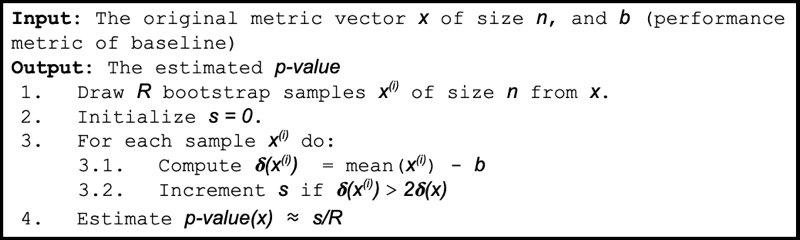}
    \caption{Pseudocode for estimating p-value \citep{berg2012empirical}}
    \label{fig:algo1}
\end{figure}

For each dataset, we generate one million bootstrap samples from the document-level F1-score vectors for our algorithm\footnote{We use \texttt{boot} package in R (\url{https://cran.r-project.org/web/packages/boot/boot.pdf}) to generate the bootstrap samples.}. Following the algorithm recommended by \cite{berg2012empirical} (Figure \ref{fig:algo1}), we estimate the p-value as the ratio of number of times our algorithm beats the baseline by twice the margin\footnote{Please refer to Section 2.2 of \cite{berg2012empirical} for a detailed discussion.} ($2\delta(x)$) on the bootstrap samples, to the total number of samples. For p-value $< 0.05$, we reject the null hypothesis.

\begin{figure}[!h]
    \centering
    \begin{subfigure}{.5\columnwidth}
        \includegraphics[scale = .23]{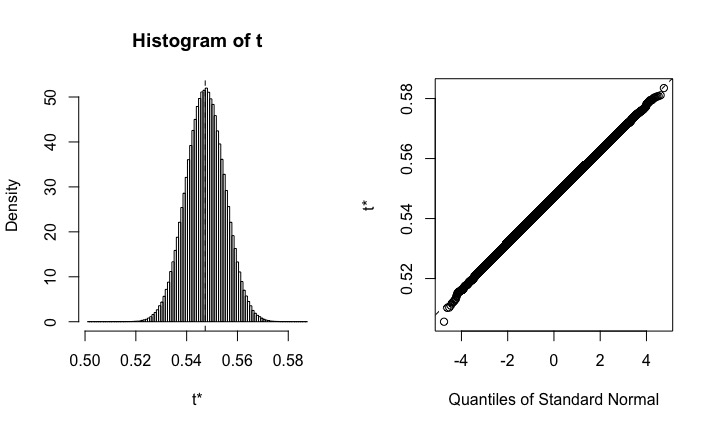}
        \caption{Hulth2003, XGB@10}
        \label{subfig:hulth-xgb@10}
    \end{subfigure}%
    \begin{subfigure}{.5\columnwidth}
        \includegraphics[scale = .23]{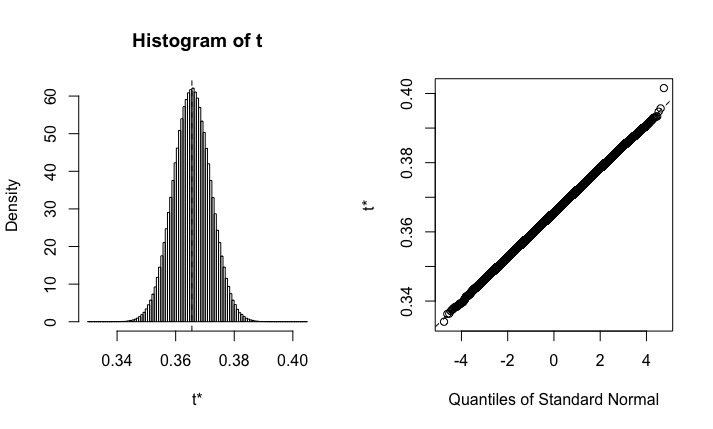}
        \caption{WWW, XGB@5}
        \label{subfig:www-xgb@5}
    \end{subfigure}
    \begin{subfigure}{.5\columnwidth}
        \includegraphics[scale = .23]{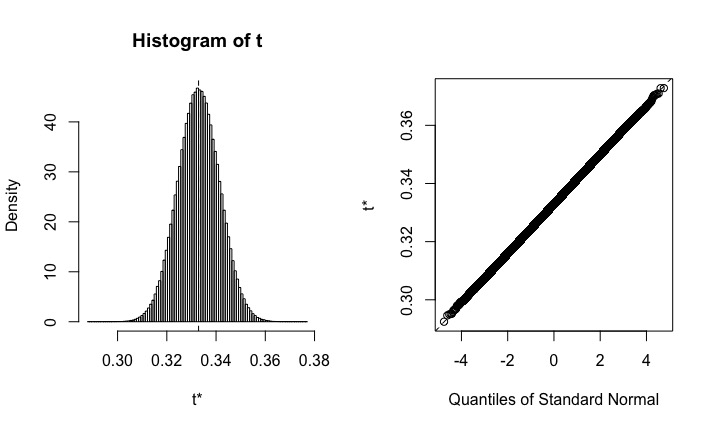}
        \caption{KDD, XGB@5}
        \label{subfig:kdd-xgb@5}
    \end{subfigure}%
    \begin{subfigure}{.5\columnwidth}
        \includegraphics[scale = .23]{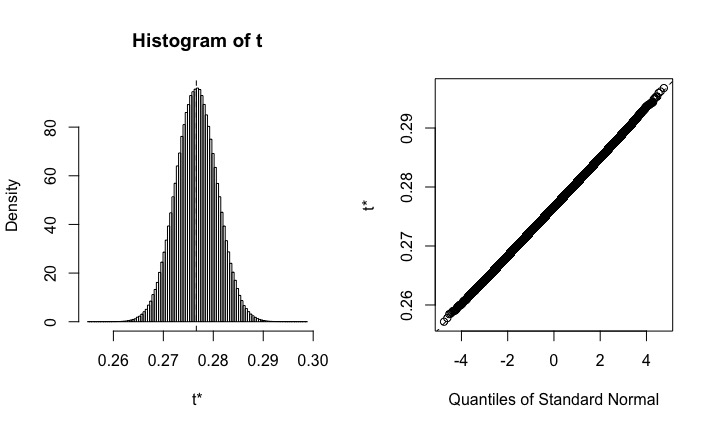}
        \caption{Krapivin2009, XGB@5}
        \label{subfig:krapi-xgb@5}
    \end{subfigure}
    \begin{subfigure}{.5\columnwidth}
        \includegraphics[scale = .23]{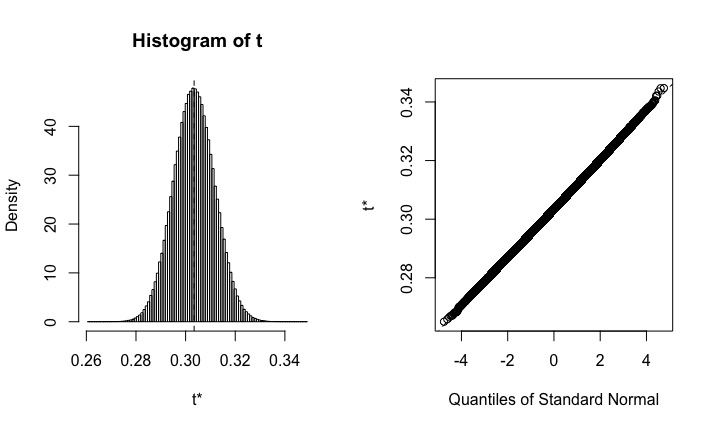}
        \caption{SemEval2010, XGB@10}
        \label{subfig:semeval-xgb@10}
    \end{subfigure}%
    \begin{subfigure}{.5\columnwidth}
        \includegraphics[scale = .23]{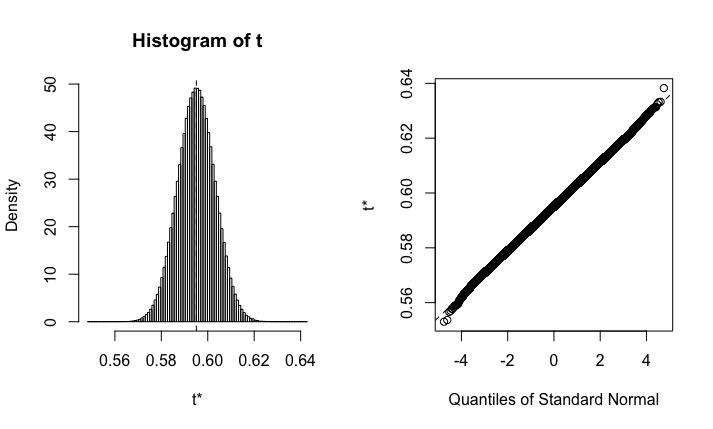}
        \caption{Marujo2012, XGB@30}
        \label{subfig:marujo-xgb@30}
    \end{subfigure}
    \caption{Distribution of average F1-scores for $R=10^6$ bootstrap samples drawn from each dataset, along with corresponding quantiles of standard normal distribution. $t$: Mean F1-score for bootstrap sample.}
    \label{Fig:distribution-statistical-significance}
\end{figure}

For all datasets except Krapivin2009, low p-value  ($< 0.05$) led to rejection of H0. This is a strong evidence that the superior performance of the proposed method is not due to chance. As evident in Table \ref{tab:krapi-res}, performance of our method is weaker than the competing method for Krapivin2009 corpus. The same is confirmed by the statistical test. We show the distribution of F1-scores for one million bootstrap samples for each dataset (for XGB model) in Figure \ref{Fig:distribution-statistical-significance}. Each plot is paired with the corresponding quantiles of standard normal distribution. Distribution of F1-scores is found to be good normal fit (Figure \ref{subfig:hulth-xgb@10}-\ref{subfig:marujo-xgb@30}) for all datasets including Krapivin2009. The mean and standard deviation for each of these distributions is shown in Table \ref{tab:mean-var}. Low standard deviation values establish consistency of the proposed method.

\begin{table}[!h]
    \scriptsize
      \centering
        \caption{Mean ($\mu$) and standard deviation ($\sigma$) of distributions shown in Figure \ref{Fig:distribution-statistical-significance}.}
        \label{tab:mean-var}
        \begin{tabular}{ c | c | c }
        \hline
        \textbf{Datasets} & \textbf{Mean} & \textbf{Standard Deviation} \\
        \hline
            Hulth2003 & 0.547325 & 0.007697 \\
            WWW & 0.365530 & 0.006432 \\
            KDD & 0.332905 & 0.008523 \\
            SemEval2010 & 0.303430 & 0.008322 \\
            Krapivin2009 & 0.276609 & 0.004156 \\
            Marujo2012 & 0.595079 & 0.008096 \\
        \hline
        \end{tabular}
    \end{table}

At this point in time, we are unable to explain consistently low performance of our method on Krapivin2009 dataset for keyphrase extraction (lower by $\approx 5\%$). Deeper investigation about the nature of Krapivin documents is pending for future.

\subsection{Keyword Extraction from Indian Language Documents}
\label{subsec:result-indian-language}
India is a country with 23 official languages, including English. According to Census of India of 2011, India has 121 major languages with more than 10000 speakers for each language\footnote{\url{http://censusindia.gov.in/2011Census/C-16_25062018_NEW.pdf}}. With such a wide variety of written and spoken languages, there is a huge collection of literature available in the country. Since digital texts are increasing day by day, automatic analysis of such documents needs to be addressed. However, due to unavailability of sophisticated NLP tools, documents written in Indian regional languages, which are grossly under-resourced, remain poorly analyzed.

We demonstrate the language-agnostic character of the proposed method by using the XGB model trained on English language documents to predict keywords from text documents written in two Indian languages. We establish the effectiveness of the proposed method in two phases. In the first phase, we choose an English document and predict the keywords. We Google translate\footnote{\url{https://translate.google.com/}. The text, stopword list, and associated codes are available at \url{https://github.com/SDuari/Supervised-Keyword-Extraction}.} the same document to Hindi and compare the keywords predicted from the translation with keywords predicted from the English document. We choose to translate an English document to Hindi over an article originally written in Hindi so that the quality of predicted Hindi keywords can be compared with the English gold-standard. In the second phase, we apply the same XGB model on five Assamese language documents. Below, we describe in detail the experiments and the observations.

The sample English text is a randomly chosen document from Marujo2012 dataset (id ``art\_and\_culture-20925876.txt"), which is translated to Hindi. We combine\footnote{We are aware that the stopword lists are not perfect, and they missed a few stopwords. However, we do not improve on the stopword list as it is out of scope for this study.} two publicly available Hindi stopwords lists\footnote{\url{https://github.com/stopwords-iso/stopwords-hi} and \url{https://www.ranks.nl/stopwords/hindi}} to create an expanded stop-list. Table \ref{tab:result_hindi} presents a detailed analysis of the results. The columns correspond to results for Hindi and English text, respectively. First row of the table lists predicted keywords using the XGB model. Based on the English gold-standard keywords list, we highlight each recalled keyword in bold. For Hindi keywords, we highlight the words whose English translation is present in the gold-standard list. Next row  presents  English translation for every Hindi keyword predicted by the model. The `-' in the translation denotes that the corresponding  word is semantically a Hindi stopword but is not included in the stop-list. Third row lists twenty keywords that are predicted from both Hindi and English versions, with the translations given in parenthesis. Out of twenty-nine total predicted keywords (last row), twenty common keywords indicate fairly good performance of the model on the Hindi document although it was trained on English corpora. We are confident that human  translated Hindi text of the English document will yield improved performance. We clarify here that the same number of Hindi and English keywords matching with the gold-standard is incidental.

\begin{table}[h!]
\centering
\caption{\label{tab:result_hindi}Keywords predicted from the original English and translated Hindi text using the pre-trained XGB model. `-' in translated keywords mean the corresponding Hindi word should be a stopword.}
\includegraphics[scale = 0.55]{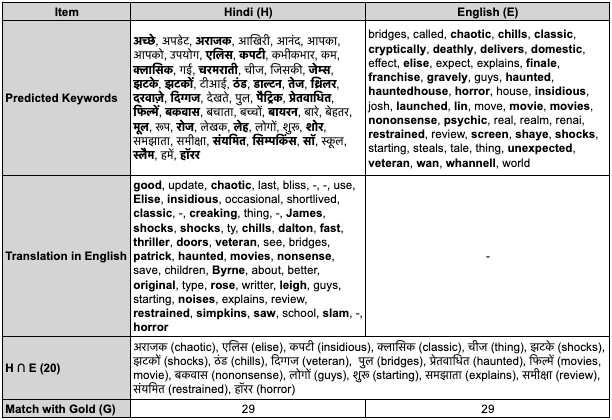}
\end{table}

In the second phase, we experiment to evaluate the model for Assamese\footnote{Assamese is the regional language of Assam, the most populous North-Eastern state of India. It is spoken by more than 15 million people, and is the mother tongue of the first author of this paper.} language texts. To perform this experiment, we collected five Assamese articles from Assamese Wikipedia\footnote{\url{https://as.wikipedia.org/}.  The collection along with the stopword list are available at \url{https://github.com/SDuari/Supervised-Keyword-Extraction}.}. The topics of the documents and the keywords predicted from each of the documents are shown in Table \ref{tab:result_assamese}. We are unable to objectively assess the performance of our method due to unavailability of gold-standard keywords for these documents. We provide English translation for the corresponding predicted keywords to enable rational assessment of the performance of our method. Keywords relevant to the topic are marked in bold. Last column shows the ratio of the relevant keywords to the total number of predicted keywords (barring ``-").

\begin{table}[!h]
\centering
\caption{\label{tab:result_assamese}Keywords predicted from the Assamese documents using the pre-trained XGB model. `-' in translated keywords mean the corresponding Assamese word is semantically a stopword. R/E: Number of relevant keywords/number of predicted keywords excluding `-'.}
\includegraphics[scale = 0.6]{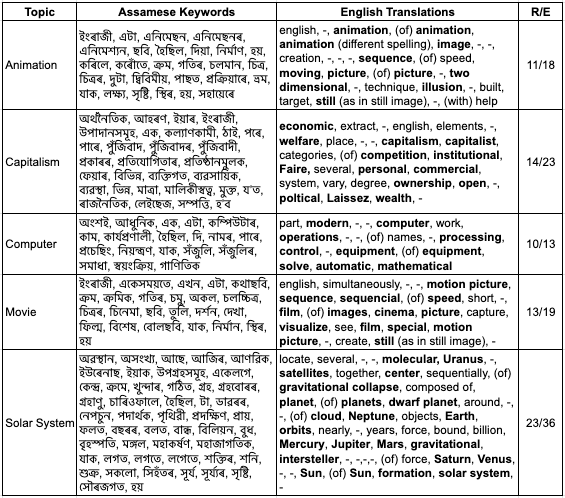}
\end{table}

Some noise is evident in the predicted keywords (e.g. `help' in \textit{Animation}, `several' in \textit{Capitalism} and \textit{Solar system}, `part in \textit{Computer}, `see' in \textit{Movie}). Interestingly, the term `english' occurs in 3/5 topics. This is because  English translations of some words are preceded by the term `english' in the Assamese text. Morphologically inflected words with different endings (translated with semantics/context in parenthesis) manifest as repetitions. For example, in \textit{Animation}, words `image', `picture', `(of) picture',  indicate to an Assamese reader that \textit{image} and \textit{picture} are keywords.

This substantiates our claim that the proposed method is applicable to any language outside the training corpus, and can perform reasonably well without using any linguistic tools. However, morphological idiosyncrasies of languages in general may have somewhat blunting effect on the potential of the proposed method. Introducing a human in the loop can quickly resolve such issues to aid automatic indexing of documents in language specific digital libraries and repositories.

\section{Conclusion}
\label{sec:conclusion}
We presented a supervised framework for automatic keyword extraction using graph-theoretic properties of words in text. The framework is domain-, collection, and language-independent. We explored six graph node properties to distinguish keywords from non-keywords - degree centrality (strength of a node), eigenvector centrality, PageRank, PositionRank, coreness, and clustering coefficient. Using training set from a mixed collection of short and long scientific texts, we trained classification models on SMOTE-balanced training set using XGBoost, Na\"ive Bayes, and bagging and boosting ensembles of Na\"ive Bayes. The induced models are then tested on four unseen collections, out of which one is from a different domain. Experimental results show that XGBoost (XGB) outperforms others in terms of F1-score, while Adabbost ensemble of Na\"ive Bayes (NB-A) closely follows. We also empirically affirm that our approach is domain- and collection-independent. Furthermore, to validate the claim of language-independence, we evaluated our models on unseen Indian language texts (Hindi and Assamese). Experimental results for keyphrase extraction show that the proposed models (XGB and NB-A) are able to outperform established keyphrase extraction models for all datasets except Krapivin2009.

Top-5 keyphrases extracted from this paper\footnote{Excluding title, keywords, Conclusion, References, footnotes, and tables and figures along with their captions.} using XGB model are - ``supervised keyword extraction", ``complex network", ``extract node properties", ``graph-based node properties", and ``keyword extraction techniques", which basically sums up the work presented here.

In future, we plan to apply the proposed approach over documents written in various Indian languages. We also intend to make our model a benchmark for cross-lingual studies, on the basis of which future keyword extraction algorithms for Indian languages could be evaluated.

\section*{References}
\bibliography{KE-bib}

\begin{thebibliography}{39}
\expandafter\ifx\csname natexlab\endcsname\relax\def\natexlab#1{#1}\fi
\providecommand{\url}[1]{\texttt{#1}}
\providecommand{\href}[2]{#2}
\providecommand{\path}[1]{#1}
\providecommand{\DOIprefix}{doi:}
\providecommand{\ArXivprefix}{arXiv:}
\providecommand{\URLprefix}{URL: }
\providecommand{\Pubmedprefix}{pmid:}
\providecommand{\doi}[1]{\href{http://dx.doi.org/#1}{\path{#1}}}
\providecommand{\Pubmed}[1]{\href{pmid:#1}{\path{#1}}}
\providecommand{\bibinfo}[2]{#2}
\ifx\xfnm\relax \def\xfnm[#1]{\unskip,\space#1}\fi
\bibitem[{Barrat et~al.(2004)Barrat, Barthelemy, Pastor-Satorras \&
  Vespignani}]{barrat2004architecture}
\bibinfo{author}{Barrat, A.}, \bibinfo{author}{Barthelemy, M.},
  \bibinfo{author}{Pastor-Satorras, R.}, \& \bibinfo{author}{Vespignani, A.}
  (\bibinfo{year}{2004}).
\newblock \bibinfo{title}{The architecture of complex weighted networks}.
\newblock {\it \bibinfo{journal}{Proceedings of the national academy of
  sciences}\/},  {\it \bibinfo{volume}{101}\/}, \bibinfo{pages}{3747--3752}.
\bibitem[{Berg-Kirkpatrick et~al.(2012)Berg-Kirkpatrick, Burkett \&
  Klein}]{berg2012empirical}
\bibinfo{author}{Berg-Kirkpatrick, T.}, \bibinfo{author}{Burkett, D.}, \&
  \bibinfo{author}{Klein, D.} (\bibinfo{year}{2012}).
\newblock \bibinfo{title}{An empirical investigation of statistical
  significance in nlp}.
\newblock In {\it \bibinfo{booktitle}{Proceedings of the 2012 Joint Conference
  on Empirical Methods in Natural Language Processing and Computational Natural
  Language Learning}\/} (pp. \bibinfo{pages}{995--1005}).
\newblock \bibinfo{organization}{Association for Computational Linguistics}.
\bibitem[{Blanco \& Lioma(2012)}]{blanco2012graph}
\bibinfo{author}{Blanco, R.}, \& \bibinfo{author}{Lioma, C.}
  (\bibinfo{year}{2012}).
\newblock \bibinfo{title}{{Graph-based Term Weighting for Information
  Retrieval}}.
\newblock {\it \bibinfo{journal}{Information Retrieval}\/},  {\it
  \bibinfo{volume}{15}\/}, \bibinfo{pages}{54--92}.
\bibitem[{Boudin(2013)}]{boudin2013comparison}
\bibinfo{author}{Boudin, F.} (\bibinfo{year}{2013}).
\newblock \bibinfo{title}{{A Comparison of Centrality Measures for Graph-based
  Keyphrase Extraction}}.
\newblock In {\it \bibinfo{booktitle}{Proceedings of the Sixth International
  Joint Conference on Natural Language Processing}\/} (pp.
  \bibinfo{pages}{834--838}).
\bibitem[{Boudin(2018)}]{boudin2018unsupervised}
\bibinfo{author}{Boudin, F.} (\bibinfo{year}{2018}).
\newblock \bibinfo{title}{{Unsupervised Keyphrase Extraction with Multipartite
  Graphs}}.
\newblock In {\it \bibinfo{booktitle}{Proceedings of the 2018 Conference of the
  North American Chapter of the Association for Computational Linguistics:
  Human Language Technologies, Volume 2 (Short Papers)}\/} (pp.
  \bibinfo{pages}{667--672}).
\newblock volume~\bibinfo{volume}{2}.
\bibitem[{Brin \& Page(1998)}]{brin1998anatomy}
\bibinfo{author}{Brin, S.}, \& \bibinfo{author}{Page, L.}
  (\bibinfo{year}{1998}).
\newblock \bibinfo{title}{{The Anatomy of a Large-scale Hypertextual Web Search
  Engine}}.
\newblock {\it \bibinfo{journal}{Computer Networks and ISDN Systems}\/},  {\it
  \bibinfo{volume}{30}\/}, \bibinfo{pages}{107--117}.
\bibitem[{Bulgarov \& Caragea(2015)}]{bulgarov2015comparison}
\bibinfo{author}{Bulgarov, F.}, \& \bibinfo{author}{Caragea, C.}
  (\bibinfo{year}{2015}).
\newblock \bibinfo{title}{{A Comparison of Supervised Keyphrase Extraction
  Models}}.
\newblock In {\it \bibinfo{booktitle}{Proceedings of the 24th International
  Conference on World Wide Web}\/} (pp. \bibinfo{pages}{13--14}).
\newblock \bibinfo{organization}{ACM}.
\bibitem[{Caragea et~al.(2014)Caragea, Bulgarov, Godea \&
  Gollapalli}]{caragea2014citation}
\bibinfo{author}{Caragea, C.}, \bibinfo{author}{Bulgarov, F.~A.},
  \bibinfo{author}{Godea, A.}, \& \bibinfo{author}{Gollapalli, S.~D.}
  (\bibinfo{year}{2014}).
\newblock \bibinfo{title}{{Citation-Enhanced Keyphrase Extraction from Research
  Papers: A Supervised Approach}}.
\newblock In {\it \bibinfo{booktitle}{Proceedings of the 2014 Conference on
  Empirical Methods in Natural Language Processing (EMNLP)}\/} (pp.
  \bibinfo{pages}{1435--1446}).
\bibitem[{Chawla et~al.(2002)Chawla, Bowyer, Hall \&
  Kegelmeyer}]{chawla2002smote}
\bibinfo{author}{Chawla, N.~V.}, \bibinfo{author}{Bowyer, K.~W.},
  \bibinfo{author}{Hall, L.~O.}, \& \bibinfo{author}{Kegelmeyer, W.~P.}
  (\bibinfo{year}{2002}).
\newblock \bibinfo{title}{{SMOTE: synthetic minority over-sampling technique}}.
\newblock {\it \bibinfo{journal}{Journal of Artificial Intelligence
  Research}\/},  {\it \bibinfo{volume}{16}\/}, \bibinfo{pages}{321--357}.
\bibitem[{Chen \& Guestrin(2016)}]{chen2016xgboost}
\bibinfo{author}{Chen, T.}, \& \bibinfo{author}{Guestrin, C.}
  (\bibinfo{year}{2016}).
\newblock \bibinfo{title}{{XGBoost: A Scalable Tree Boosting System}}.
\newblock In {\it \bibinfo{booktitle}{Proceedings of the 22nd ACM SIGKDD
  International Conference on Knowledge Discovery and Data Mining}\/} (pp.
  \bibinfo{pages}{785--794}).
\newblock \bibinfo{organization}{ACM}.
\bibitem[{Chuang et~al.(2012)Chuang, Manning \& Heer}]{chuang2012without}
\bibinfo{author}{Chuang, J.}, \bibinfo{author}{Manning, C.~D.}, \&
  \bibinfo{author}{Heer, J.} (\bibinfo{year}{2012}).
\newblock \bibinfo{title}{{“Without the clutter of unimportant words”:
  Descriptive keyphrases for text visualization}}.
\newblock {\it \bibinfo{journal}{ACM Transactions on Computer-Human Interaction
  (TOCHI)}\/},  {\it \bibinfo{volume}{19}\/}, \bibinfo{pages}{19}.
\bibitem[{Dror et~al.(2018)Dror, Baumer, Shlomov \&
  Reichart}]{dror2018hitchhiker}
\bibinfo{author}{Dror, R.}, \bibinfo{author}{Baumer, G.},
  \bibinfo{author}{Shlomov, S.}, \& \bibinfo{author}{Reichart, R.}
  (\bibinfo{year}{2018}).
\newblock \bibinfo{title}{The hitchhiker’s guide to testing statistical
  significance in natural language processing}.
\newblock In {\it \bibinfo{booktitle}{Proceedings of the 56th Annual Meeting of
  the Association for Computational Linguistics (Volume 1: Long Papers)}\/}
  (pp. \bibinfo{pages}{1383--1392}).
\bibitem[{Duari \& Bhatnagar(2019)}]{duari2019scake}
\bibinfo{author}{Duari, S.}, \& \bibinfo{author}{Bhatnagar, V.}
  (\bibinfo{year}{2019}).
\newblock \bibinfo{title}{{sCAKE: Semantic Connectivity Aware Keyword
  Extraction}}.
\newblock {\it \bibinfo{journal}{Information Sciences}\/},  {\it
  \bibinfo{volume}{477}\/}, \bibinfo{pages}{100--117}.
\bibitem[{Florescu \& Caragea(2017)}]{florescu2017position}
\bibinfo{author}{Florescu, C.}, \& \bibinfo{author}{Caragea, C.}
  (\bibinfo{year}{2017}).
\newblock \bibinfo{title}{{A Position-Biased PageRank Algorithm for Keyphrase
  Extraction}}.
\newblock In {\it \bibinfo{booktitle}{Proceedings of Thirty-First AAAI
  Conference on Artificial Intelligence}\/} (pp. \bibinfo{pages}{4923--4924}).
\bibitem[{Gollapalli et~al.(2017)Gollapalli, Li \&
  Yang}]{gollapalli2017incorporating}
\bibinfo{author}{Gollapalli, S.~D.}, \bibinfo{author}{Li, X.-L.}, \&
  \bibinfo{author}{Yang, P.} (\bibinfo{year}{2017}).
\newblock \bibinfo{title}{{Incorporating Expert Knowledge into Keyphrase
  Extraction}}.
\newblock In {\it \bibinfo{booktitle}{Thirty-First AAAI Conference on
  Artificial Intelligence}\/} (pp. \bibinfo{pages}{3180--3187}).
\bibitem[{Herrera \& Pury(2008)}]{herrera2008statistical}
\bibinfo{author}{Herrera, J.~P.}, \& \bibinfo{author}{Pury, P.~A.}
  (\bibinfo{year}{2008}).
\newblock \bibinfo{title}{{{Statistical Keyword Detection in Literary
  Corpora}}}.
\newblock {\it \bibinfo{journal}{The European Physical Journal B}\/},  {\it
  \bibinfo{volume}{63}\/}, \bibinfo{pages}{135--146}.
\bibitem[{Hulth(2003)}]{hulth2003improved}
\bibinfo{author}{Hulth, A.} (\bibinfo{year}{2003}).
\newblock \bibinfo{title}{{{Improved Automatic Keyword Extraction given more
  Linguistic Knowledge}}}.
\newblock In {\it \bibinfo{booktitle}{Proceedings of the 2003 conference on
  Empirical methods in natural language processing}\/} (pp.
  \bibinfo{pages}{216--223}).
\newblock \bibinfo{organization}{Association for Computational Linguistics}.
\bibitem[{Kim et~al.(2010)Kim, Medelyan, Kan \& Baldwin}]{kim2010semeval}
\bibinfo{author}{Kim, S.~N.}, \bibinfo{author}{Medelyan, O.},
  \bibinfo{author}{Kan, M.-Y.}, \& \bibinfo{author}{Baldwin, T.}
  (\bibinfo{year}{2010}).
\newblock \bibinfo{title}{{Semeval-2010 Task 5: Automatic Keyphrase Extraction
  from Scientific Articles}}.
\newblock In {\it \bibinfo{booktitle}{Proceedings of the 5th International
  Workshop on Semantic Evaluation}\/} (pp. \bibinfo{pages}{21--26}).
\newblock \bibinfo{organization}{Association for Computational Linguistic}.
\bibitem[{Krapivin et~al.(2009)Krapivin, Autaeu \&
  Marchese}]{krapivin2009large}
\bibinfo{author}{Krapivin, M.}, \bibinfo{author}{Autaeu, A.}, \&
  \bibinfo{author}{Marchese, M.} (\bibinfo{year}{2009}).
\newblock \bibinfo{title}{{Large Dataset for Keyphrases Extraction}}.
\newblock {\it \bibinfo{journal}{Technical Report DISI-09-055}\/}, .
\bibitem[{Litvak et~al.(2011)Litvak, Last, Aizenman, Gobits \&
  Kandel}]{litvak2011degext}
\bibinfo{author}{Litvak, M.}, \bibinfo{author}{Last, M.},
  \bibinfo{author}{Aizenman, H.}, \bibinfo{author}{Gobits, I.}, \&
  \bibinfo{author}{Kandel, A.} (\bibinfo{year}{2011}).
\newblock \bibinfo{title}{{DegExt—A Language-independent Graph-based
  Keyphrase Extractor}}.
\newblock In {\it \bibinfo{booktitle}{Advances in Intelligent Web
  Mastering--3}\/} (pp. \bibinfo{pages}{121--130}).
\newblock \bibinfo{publisher}{Springer}.
\bibitem[{Lopez \& Romary(2010{\natexlab{a}})}]{lopez2010grisp}
\bibinfo{author}{Lopez, P.}, \& \bibinfo{author}{Romary, L.}
  (\bibinfo{year}{2010}{\natexlab{a}}).
\newblock \bibinfo{title}{{GRISP: A Massive Multilingual Terminological
  Database for Scientific and Technical Domains}}.
\newblock In {\it \bibinfo{booktitle}{Proceedings of the Seventh conference on
  International Language Resources and Evaluation (LREC'10)}\/}.
\bibitem[{Lopez \& Romary(2010{\natexlab{b}})}]{lopez2010humb}
\bibinfo{author}{Lopez, P.}, \& \bibinfo{author}{Romary, L.}
  (\bibinfo{year}{2010}{\natexlab{b}}).
\newblock \bibinfo{title}{{HUMB: Automatic key term extraction from scientific
  articles in GROBID}}.
\newblock In {\it \bibinfo{booktitle}{Proceedings of the 5th international
  workshop on semantic evaluation}\/} (pp. \bibinfo{pages}{248--251}).
\newblock \bibinfo{organization}{Association for Computational Linguistics}.
\bibitem[{Luhn(1957)}]{luhn1957statistical}
\bibinfo{author}{Luhn, H.~P.} (\bibinfo{year}{1957}).
\newblock \bibinfo{title}{{A Statistical Approach to Mechanized Encoding and
  Searching of Literary Information}}.
\newblock {\it \bibinfo{journal}{IBM Journal of R\&D}\/},  {\it
  \bibinfo{volume}{1}\/}, \bibinfo{pages}{309--317}.
\bibitem[{Marujo et~al.(2012)Marujo, Gershman, Carbonell, Frederking \&
  Neto}]{marujo2012supervised}
\bibinfo{author}{Marujo, L.}, \bibinfo{author}{Gershman, A.},
  \bibinfo{author}{Carbonell, J.}, \bibinfo{author}{Frederking, R.}, \&
  \bibinfo{author}{Neto, J.~P.} (\bibinfo{year}{2012}).
\newblock \bibinfo{title}{{Supervised Topical Key Phrase Extraction of News
  Stories using Crowdsourcing, Light Filtering and Co-reference
  Normalization}}.
\newblock In {\it \bibinfo{booktitle}{Proceedings of the Eighth International
  Conference on Language Resources and Evaluation (LREC-2012)}\/}.
\bibitem[{Medelyan et~al.(2009)Medelyan, Frank \& Witten}]{medelyan2009human}
\bibinfo{author}{Medelyan, O.}, \bibinfo{author}{Frank, E.}, \&
  \bibinfo{author}{Witten, I.~H.} (\bibinfo{year}{2009}).
\newblock \bibinfo{title}{{Human-competitive tagging using automatic keyphrase
  extraction}}.
\newblock In {\it \bibinfo{booktitle}{Proceedings of the 2009 Conference on
  Empirical Methods in Natural Language Processing: Volume 3}\/} (pp.
  \bibinfo{pages}{1318--1327}).
\newblock \bibinfo{organization}{Association for Computational Linguistics}
  volume~\bibinfo{volume}{3}.
\bibitem[{Mihalcea \& Tarau(2004)}]{mihalcea2004textrank}
\bibinfo{author}{Mihalcea, R.}, \& \bibinfo{author}{Tarau, P.}
  (\bibinfo{year}{2004}).
\newblock \bibinfo{title}{{{Textrank: Bringing order into text}}}.
\newblock In {\it \bibinfo{booktitle}{Proceedings of the 2004 conference on
  empirical methods in natural language processing}\/}.
\bibitem[{Mothe et~al.(2018)Mothe, Ramiandrisoa \&
  Rasolomanana}]{mothe2018automatic}
\bibinfo{author}{Mothe, J.}, \bibinfo{author}{Ramiandrisoa, F.}, \&
  \bibinfo{author}{Rasolomanana, M.} (\bibinfo{year}{2018}).
\newblock \bibinfo{title}{Automatic keyphrase extraction using graph-based
  methods}.
\newblock In {\it \bibinfo{booktitle}{Proceedings of the 33rd Annual ACM
  Symposium on Applied Computing}\/} (pp. \bibinfo{pages}{728--730}).
\newblock \bibinfo{organization}{ACM}.
\bibitem[{Nguyen \& Kan(2007)}]{nguyen2007keyphrase}
\bibinfo{author}{Nguyen, T.~D.}, \& \bibinfo{author}{Kan, M.-Y.}
  (\bibinfo{year}{2007}).
\newblock \bibinfo{title}{{Keyphrase Extraction in Scientific Publications}}.
\newblock In {\it \bibinfo{booktitle}{International conference on Asian digital
  libraries}\/} (pp. \bibinfo{pages}{317--326}).
\newblock \bibinfo{organization}{Springer}.
\bibitem[{Ortuno et~al.(2002)Ortuno, Carpena, Bernaola-Galv{\'a}n, Mu{\~n}oz \&
  Somoza}]{ortuno2002keyword}
\bibinfo{author}{Ortuno, M.}, \bibinfo{author}{Carpena, P.},
  \bibinfo{author}{Bernaola-Galv{\'a}n, P.}, \bibinfo{author}{Mu{\~n}oz, E.},
  \& \bibinfo{author}{Somoza, A.} (\bibinfo{year}{2002}).
\newblock \bibinfo{title}{{Keyword Detection in Natural Languages and DNA}}.
\newblock {\it \bibinfo{journal}{EPL (Europhysics Letters)}\/},  {\it
  \bibinfo{volume}{57}\/}, \bibinfo{pages}{759}.
\bibitem[{Papagiannopoulou \& Tsoumakas(2018)}]{papagiannopoulou2018local}
\bibinfo{author}{Papagiannopoulou, E.}, \& \bibinfo{author}{Tsoumakas, G.}
  (\bibinfo{year}{2018}).
\newblock \bibinfo{title}{{Local word vectors guiding keyphrase extraction}}.
\newblock {\it \bibinfo{journal}{Information Processing \& Management}\/},
  {\it \bibinfo{volume}{54}\/}, \bibinfo{pages}{888--902}.
\bibitem[{Rousseau \& Vazirgiannis(2015)}]{rousseau2015main}
\bibinfo{author}{Rousseau, F.}, \& \bibinfo{author}{Vazirgiannis, M.}
  (\bibinfo{year}{2015}).
\newblock \bibinfo{title}{{{Main core retention on graph-of-words for
  single-document keyword extraction}}}.
\newblock In {\it \bibinfo{booktitle}{European Conference on Information
  Retrieval}\/} (pp. \bibinfo{pages}{382--393}).
\newblock \bibinfo{organization}{Springer}.
\bibitem[{Seidman(1983)}]{seidman1983network}
\bibinfo{author}{Seidman, S.~B.} (\bibinfo{year}{1983}).
\newblock \bibinfo{title}{{Network Structure and Minimum Degree}}.
\newblock {\it \bibinfo{journal}{Social Networks}\/},  {\it
  \bibinfo{volume}{5}\/}, \bibinfo{pages}{269--287}.
\bibitem[{Sterckx et~al.(2016)Sterckx, Demeester, Develder \&
  Caragea}]{sterckx2016supervised}
\bibinfo{author}{Sterckx, L.}, \bibinfo{author}{Demeester, T.},
  \bibinfo{author}{Develder, C.}, \& \bibinfo{author}{Caragea, C.}
  (\bibinfo{year}{2016}).
\newblock \bibinfo{title}{{Supervised Keyphrase Extraction as Positive
  Unlabeled Learning}}.
\newblock In {\it \bibinfo{booktitle}{Proceedings of the 2016 Conference on
  Empirical Methods in Natural Language Processing}\/} (pp.
  \bibinfo{pages}{1--6}).
\bibitem[{Tixier et~al.(2016)Tixier, Malliaros \&
  Vazirgiannis}]{tixier2016graph}
\bibinfo{author}{Tixier, A.}, \bibinfo{author}{Malliaros, F.}, \&
  \bibinfo{author}{Vazirgiannis, M.} (\bibinfo{year}{2016}).
\newblock \bibinfo{title}{{A Graph Degeneracy-based Approach to Keyword
  Extraction}}.
\newblock In {\it \bibinfo{booktitle}{Proceedings of the 2016 Conference on
  Empirical Methods in Natural Language Processing6}\/} (pp.
  \bibinfo{pages}{1860--1870}).
\bibitem[{Turney(1999)}]{turney1999learning}
\bibinfo{author}{Turney, P.~D.} (\bibinfo{year}{1999}).
\newblock \bibinfo{title}{{Learning to Extract Keyphrases from Text}}.
\newblock {\it \bibinfo{journal}{Technical Report, National Research Council of
  Canada}\/}, .
\bibitem[{Witten et~al.(1999)Witten, Paynter, Frank, Gutwin \&
  Nevill-Manning}]{witten1999kea}
\bibinfo{author}{Witten, I.~H.}, \bibinfo{author}{Paynter, G.~W.},
  \bibinfo{author}{Frank, E.}, \bibinfo{author}{Gutwin, C.}, \&
  \bibinfo{author}{Nevill-Manning, C.~G.} (\bibinfo{year}{1999}).
\newblock \bibinfo{title}{{KEA: Practical Automatic Keyphrase Extraction}}.
\newblock In {\it \bibinfo{booktitle}{Proceedings of the Fourth ACM Conference
  on Digital Libraries}\/} (pp. \bibinfo{pages}{254--255}).
\newblock \bibinfo{organization}{ACM}.
\bibitem[{Zaki et~al.(2014)Zaki, Meira~Jr \& Meira}]{zaki2014data}
\bibinfo{author}{Zaki, M.~J.}, \bibinfo{author}{Meira~Jr, W.}, \&
  \bibinfo{author}{Meira, W.} (\bibinfo{year}{2014}).
\newblock {\it \bibinfo{title}{{Data Mining and Analysis: Fundamental Concepts
  and Algorithms}}\/}.
\newblock \bibinfo{publisher}{Cambridge University Press}.
\bibitem[{Zhang(2008)}]{zhang2008automatic}
\bibinfo{author}{Zhang, C.} (\bibinfo{year}{2008}).
\newblock \bibinfo{title}{{Automatic Keyword Extraction from Documents using
  Conditional Random Fields}}.
\newblock {\it \bibinfo{journal}{Journal of Computational Information
  Systems}\/},  {\it \bibinfo{volume}{4}\/}, \bibinfo{pages}{1169--1180}.
\bibitem[{Zhang et~al.(2017)Zhang, Chang, Liu, Gollapalli, Li \&
  Xiao}]{zhang2017mike}
\bibinfo{author}{Zhang, Y.}, \bibinfo{author}{Chang, Y.}, \bibinfo{author}{Liu,
  X.}, \bibinfo{author}{Gollapalli, S.~D.}, \bibinfo{author}{Li, X.}, \&
  \bibinfo{author}{Xiao, C.} (\bibinfo{year}{2017}).
\newblock \bibinfo{title}{{MIKE: keyphrase extraction by integrating
  multidimensional information}}.
\newblock In {\it \bibinfo{booktitle}{Proceedings of the 2017 ACM on Conference
  on Information and Knowledge Management}\/} (pp.
  \bibinfo{pages}{1349--1358}).
\newblock \bibinfo{organization}{ACM}.

\end{thebibliography}

\end{document}